%% file: main.tex
\documentclass[%
reprint,
amsmath,amssymb,
aip,
jcp,
]{revtex4-2}

\setcitestyle{super,sort&compress}

\usepackage[english]{babel}
\usepackage{graphicx}
\usepackage{float}
\usepackage{xcolor}
\usepackage{dcolumn}
\usepackage{bm}
\usepackage{siunitx}
\usepackage[nolist]{acronym}
\usepackage[version=4]{mhchem}
\usepackage{natmove}
\usepackage{xr} 
\usepackage[colorlinks=true,linkcolor=blue,citecolor=blue,urlcolor=blue]{hyperref}
\usepackage{cleveref}

\DeclareSIUnit\elementarycharge{\text{\ensuremath{e}}}
\DeclareSIUnit\angstrom{\text{Å}}
\Crefname{equation}{Eq.}{Eqs.}
\crefname{equation}{Eq.}{Eqs.}
\creflabelformat{equation}{#2#1#3}
\Crefname{figure}{Fig.}{Figs.}
\crefname{figure}{Fig.}{Figs.}
\Crefname{table}{Tab.}{Tabs.}
\crefname{table}{Tab.}{Tabs.}

\newcommand{\kb}[1]{k_\mathrm{B}}
\newcommand{\eg}[1]{\textit{e.g.}\@}
\newcommand{\ie}[1]{\textit{i.e.}\@}
\newcommand{\cf}[1]{\textit{cf.}\@}
\newcommand{\dd}{\mathrm{d}}

\makeatletter
\AtBeginDocument
 {
   \def\ltx@label#1{\cref@label{#1}}
    \def\label@in@display@noarg#1{\cref@old@label@in@display{#1}}
\def\label@in@mmeasure@noarg#1{%
    \begingroup%
      \measuring@false%
      \cref@old@label@in@display{#1}
    \endgroup}%
 } %
\makeatother

\makeatletter
\def\@email#1#2{%
 \endgroup
 \patchcmd{\titleblock@produce}
  {\frontmatter@RRAPformat}
  {\frontmatter@RRAPformat{\produce@RRAP{*#1\href{mailto:#2}{#2}}}\frontmatter@RRAPformat}
  {}{}
}%

\begin{document}
\preprint{APS/123-QED}

\title{Electrochemical Molecular Dynamics within the Finite-Field Framework}

\author{Krystof Brezina$^*$}
\email[Corresponding author: ]{krystof.brezina@mpsd.mpg.de}
\affiliation{Max Planck Institute for the Structure and Dynamics of Matter, Luruper Chaussee 149, 22761 Hamburg, Germany}
\author{Ludwig Ahrens-Iwers}
\affiliation{Max Planck Institute for the Structure and Dynamics of Matter, Luruper Chaussee 149, 22761 Hamburg, Germany}
\author{Mariana Rossi$^*$}
\email[Corresponding author: ]{mr2102@cam.ac.uk}
\affiliation{Max Planck Institute for the Structure and Dynamics of Matter, Luruper Chaussee 149, 22761 Hamburg, Germany}
\affiliation{Yusuf Hamied Department of Chemistry, University of Cambridge, Lensfield Road, Cambridge CB2 1EW, U.K.}

\date{\today}

\begin{abstract}
Electrochemical processes at charged interfaces lie at the heart of modern technologies and their continued improvement depends on insights into the relevant atomistic processes.
However, the necessary tools to impose constant-voltage conditions in molecular dynamics (MD) simulations typically directly target the electrode charge, making their extension to methods that go beyond an empirical description of metals more cumbersome.
In this work, we propose an MD protocol to realize a thermodynamic ensemble in which voltage control is attained, but which is  formulated purely within the framework of finite-field simulations in 3D periodic boundary conditions. 
We show a derivation and implementation of such a control device, which we coin an electrostat, in the \textsc{Electrode} package within \textsc{Lammps}. 
Our simulations systematically benchmark the derived theory against other established approaches, for example, showing its equivalence to a potentiostat,  and quantify the importance of properly accounting for ensemble fluctuations.
We find that for systems of several thousands of atoms, one obtains the same equilibrium properties with the electrostat as with a simpler constant-field simulation.
The difference between the two methods only manifests itself at smaller length scales, thereby establishing practical limits for the inclusion of voltage fluctuations in atomistic simulations of electrochemical systems. 
The strength of this framework also lies in its simplicity of extension to first-principles \textit{ab initio} and reactive machine-learning potentials.
\end{abstract}

\maketitle

\input{acronyms.tex}

\section{\label{sec:introduction} Introduction}

Electrochemical processes are at the core of numerous critical sectors, including energy conversion and storage, metallurgy, sensing, or chemical production~\cite{chen_materials_2016,little_introduction_2018}. 
A detailed understanding of these processes is vital for developing and fine-tuning new efficient technologies.
Such an understanding requires atomistic level insight into electrochemical systems under \textit{operando} conditions, which is currently often insufficient or lacking, as both experimental imaging methods~\cite{domke_monitoring_2024} and accurate theoretical simulations~\cite{wang_theory_2021} are still at an early, albeit active stage of development.

On the theoretical side, efficient first-principles \acf{md} simulations of electrochemical systems face two main challenges.
First, the simulated system must be held at a controlled electrode potential. This is fundamentally an electronically grand-canonical problem, and addressing it within standard electronic structure methods can introduce significant computational overhead and implementation hurdles~\cite{bonnet_first-principles_2012, beinlich_controlled_2023}.
Second, electrochemical processes involve inherently heterogeneous interfaces, whose realistic description requires simulations spanning prohibitively large length and time scales.
Even for idealized electrode surfaces, the timescales for charging the system and forming an \ac{edl} in nanocapacitors reach several nanoseconds or more~\cite{pean_dynamics_2014, vatamanu_capacitive_2015}, and the need to model realistic surface reconstructions~\cite{koper_structure_2011} further exacerbates the problem.

Potential control in \ac{md} simulations is more established in the field of empirical modeling, where both issues mentioned above are addressed at the cost of forfeiting the explicit electronic structure.
Central to this field is the \acf{cpm} introduced by Siepmann and Sprik\cite{siepmann_influence_1995} and Reed \textit{et al.}~\cite{reed_electrochemical_2007}, which relies on a charge-optimization scheme to approximate perfectly conducting electrodes under fixed-potential, closed circuit conditions.
It can be straightforwardly reformulated into its open-circuit counterpart, the \acf{ccm}~\cite{ahrens-iwers_electrode_2022, tee_constant_2023,hormannConvergingDivergentPaths2024,bergmann_machine_2025}.
These (and other) electrode-based methods are widely available in established molecular modeling software, such as in the \textsc{Electrode} package\cite{ahrens-iwers_electrode_2022} in \textsc{Lammps}\cite{thompson_lammps_2022} or in MetalWalls\cite{marin-lafleche_metalwalls_2020, coretti_metalwalls_2022} and are routinely used to study  dielectric properties, dynamics and nanoscale structuring of liquid/metal interfaces.

More recently, it was discussed~\cite{bonnet_first-principles_2012, deisenbeck_dielectric_2021} that when a fixed potential is imposed, its value is only strictly constant in the thermodynamic limit. 
In finite-size systems and in accordance with the thermodynamical ensemble~\cite{landau_electrodynamics_1984}, the potential must exhibit fluctuations following the Johnson--Nyquist (fluctuation-dissipation) theorem~\cite{blundell_concepts_2017, limmer_charge_2013}.
Such considerations led to the idea of a potentiostat: a control device, similar in spirit to \ac{md} thermostats~\cite{bussi_accurate_2007} or barostats~\cite{bernetti_pressure_2020}, that ensures potential control with correct thermodynamics by coupling to a reservoir that acts as a computational voltage source.
Combined with temperature control approaches, potentiostats for molecular dynamics protocols have been formulated using an extended Lagrangian by Bonnet \textit{et al.}~\cite{bonnet_first-principles_2012} or using stochastic propagation by Deißenbeck \textit{et al.}~\cite{deisenbeck_dielectric_2021}.
Nevertheless, the extension of potentiostat methods to the first-principles framework to enable the study of interfacial electronic properties, spillover effects or chemical reactions, requires artificial electrode setups that introduce further interfaces in the system, in order to treat a 2D-periodic, and possibly charged system, as discussed in the pioneering work by Deißenbeck and coworkers~\cite{deisenbeck_dielectric_2023} or in a recent implementation of the potentiostat~\cite{zhangFlexibleGeneralizedConstantPotential2026} in the i-PI software~\cite{litmanIPI30Flexible2024}.

A viable alternative to couple charged interfaces with first-principles methods relies on the modern theory of polarization~\cite{resta_quantum-mechanical_1998, spaldin_beginners_2012}, which allows \ac{md} simulations of systems under finite-field electrical boundary conditions~\cite{zhang_finite_2016}, which control the macroscopic electric field $\mathbf{E}$\cite{umari_ab_2002}, or the displacement field $\mathbf{D}$\cite{stengel_electric_2009}. 
In particular, numerous studies have focused on applying these methods to water and interfaces present in electrochemical setups.  
These studies involve first-principles simulations of dielectric properties~\cite{stocco_electric-field_2025, zhang_computing-dielectric_2016}, electrofreezing~\cite{cassone_electrofreezing_2024, stocco_electric-field_2025}, molecular dissociation~\cite{cassone_nuclear_2020, litman_entropy_2025}, vibrational spectroscopy~\cite{cassone_ab_2019}, as well as characterization of interfacial phenomena~\cite{wang_ion-modulated_2026, andersson_electronic_2025} and machine learning of  interfacial charge distributions~\cite{grisafi_predicting_2023, grisafi_accelerating_2024}.

The key difference between the finite-field methods and the electrode-based methods is that, in the former case, no charge distributions are enforced explicitly, but rather are obtained directly by the polarization induced by the imposed external fields.
Such methods are formulated under 3D \acp{pbc}, are computationally significantly more efficient~\cite{dufils_simulating_2019, tee_fully_2022}, readily compatible with first-principles electronic structure infrastructures~\cite{umari_ab_2002, andersson_electronic_2025} and show much promise to be connected to field-aware machine-learning models of first-principles potential energy surfaces.~\cite{stocco_electric-field_2025, batatia_mace-polar-1_2026, feng_machine_2025, wang_ion-modulated_2026}
The constant-$E$ and constant-$D$ ensembles are directly connected to the closed-circuit \ac{cpm} and open-circuit \ac{ccm} ensembles, respectively, and, therefore, can be used as a means of imposing electrochemical conditions.
This fact was recognized in the seminal paper by Stengel \textit{et al.}~\cite{stengel_electric_2009}, discussed in the theoretical works by Sprik~\cite{sprik_finite_2018, zhang_finite_2016} and later demonstrated through numerical simulations by Dufils \textit{et al.}~\cite{dufils_simulating_2019,dufils_computational_2021}.
We point out that that a known limitation of finite-field methods in a first-principles context is the onset of the Zener breakdown, which places an upper bound on the applicable field magnitude.~\cite{souza_first-principles_2002}.

In this paper, we propose a framework that joins advantages from two sides: we impose the correct ensemble fluctuations just like in the potentiostat formulation, but exploit the computational advantages of finite-field methods. 
To achieve this, we formulate an electrostat, that imposes potential control by means of a fluctuating $\mathbf{E}$-field. We also describe an implementation of the method coupled to force-field simulations, together with benchmark results and analysis.

The paper is organized as follows.
In Section~\ref{sec:Theory}, we discuss the theory of the electrostat and present an overview of its implementation in the \textsc{Electrode} package of \textsc{Lammps}.
In Section~\ref{sec:Results and Discussion} we show and discuss several important benchmarks and results obtained with the electrostat.
These provide numerical proofs of the concepts discussed in Sections~\ref{sec:Theory}, show applications to solvated ion association energetics and diffusive dynamics of an ionic liquid under potential biases and compare the electrostat with the constant-$E$ method to understand the impact of correctly describing potential fluctuations.
Finally, Section~\ref{sec:Conclusion} summarizes and concludes the paper.

\section{\label{sec:Theory} Theory and Methodology}

\subsection{Stochastic sampling of the constant-$E$ ensemble with correct fluctuations}

\begin{figure*}[tb!]
    \centering
    \includegraphics[width=\linewidth]{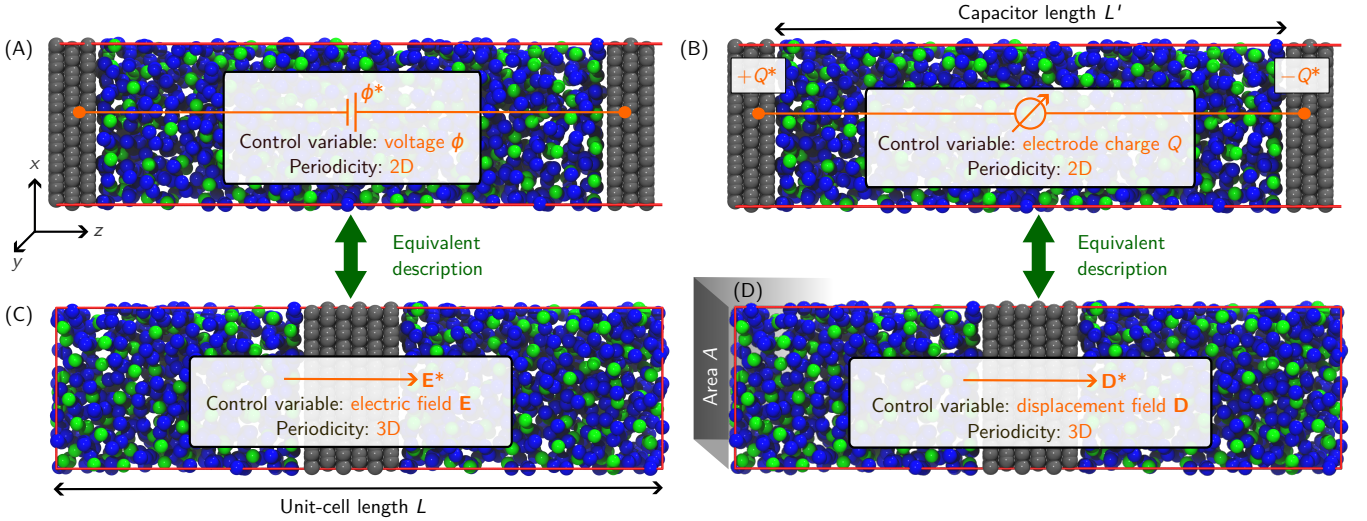}
    \caption{
    The possible simulation approaches for a nanoscopic electrochemical cell consisting of a coarse-grained ionic liquid electrolyte (butylmethylimidazolium hexafluorophosphate, [BMIM$^+$][PF$_6^-$], with the cation shown in blue and the anion in green) and planar graphite electrodes (carbon atoms are shown in gray).
    Boundaries, across which \acp{pbc} are applied, are shown in red.
    The lateral ($z$) length of the unit cell is labeled $L$ and capacitor (plate-to-plate) length $L'$.
    In panel A, we show the closed-circuit set up consistent with the \ac{cpm}.
    Here, the unit cell has 2D \ac{pbc} in the ($x, y$)-directions with the electrodes facing a vacuum on the outside edges of the unit cell and the control variable is the voltage fixed to a chosen value $\phi^*$.
    Allowing for thermodynamically consistent fluctuations of the voltage corresponds to the stochastic potentiostat~\cite{deisenbeck_dielectric_2021} in the same configuration.
    Panel B shows the open-circuit version of the system in A, where the total charge per electrode is constrained to a value of $\pm Q^*$ by the \ac{ccm}.
    Panel C shows a representation of the same cell under 3D \ac{pbc} and shifted in the $z$ direction by $L/2$: here, the two electrodes have merged into a single one and the electrode--vacuum interface is no longer present.
    Such a system can be treated with the constant-$E$ method that imposes a finite field $E^*$: when the magnitude of the field is chosen to be $-\phi^* / L$, the \ac{cpm} (A) and constant-$E$ method (C) yield an equivalent description, as marked by the green arrow.
    Allowing for fluctuations of $E$ leads to the electrostat.
    Panel D shows the 3D-\ac{pbc} finite-field counterpart of the \ac{ccm} setup (B) with a constant displacement field $D^*$ which is equivalent to \ac{ccm} with the field magnitude set to $Q^* / A$ (see Appendix~\ref{app:theory-equivalence} for details.)
    }
    \label{fig:cell-definition}
\end{figure*}

In Figure~\ref{fig:cell-definition}, we show representative examples of relevant nanocapacitor geometries together with a graphical summary of how electrochemical ensembles can be imposed and which equivalences apply among them.
In the rest of the paper, we assume that the positions of the electrode atoms are fixed, but this assumption can also be relaxed, see Appendix~\ref{app:theory-equivalence}.
Crucially for this work, we rely on the equivalence between the 2D-periodic nanocapacitor kept under constant voltage $\phi^*$ through the \ac{cpm} and the 3D-periodic nanocapacitor kept under a constant field 
\begin{equation}
    E^* = -\frac{\phi^*}{L} \label{eq:e-star}
\end{equation} 
with $L$ the unit cell length (see \cref{fig:cell-definition}A and \ref{fig:cell-definition}C). 
Throughout this paper, we will use asterisks ${}^*$ to denote the target value of the control variables.
Here and in the rest of the paper, we assume all  fields to be $z$-oriented, and for ease of notation thus consider $E$ to contain only a non-zero $z$ component.
Additionally, a similar equivalence holds for the open-circuit case, where the quantity constrained via the \ac{ccm} in the 2D-periodic geometry is the total electrode charge $Q^*$ and in the 3D-periodic geometry the displacement field $D^* = Q^*/A$, where $A$ is the electrode area per unit cell (see Figure~\ref{fig:cell-definition}D).
An extended discussion of these known ensemble equivalences based on a more rigorous mathematical basis is presented in Appendix~\ref{app:theory-equivalence}. 

We now proceed to generalize the constant-$E$ ensemble to accommodate thermal noise in the circuit which leads to the electrostat.
The Hamiltonian of the constant-$E$ system $H_E$ is defined (in SI units) as a  Legendre transform of the constant-$D$ Hamiltonian~\cite{landau_electrodynamics_1984, stengel_electric_2009, sprik_finite_2018} 
\begin{equation}
\begin{split}
    & H_E\left(\mathbf{r}, \mathbf{p}; E^*\right) \\
    &= \inf_D \min_\psi
    \underbrace{\left\{ H_0(\mathbf{r}, \mathbf{p}; \psi)
    + \frac{\Omega}{2\varepsilon_0}\left[D - P(\mathbf{r}; \psi)\right]^2
    - \Omega D E^* \right\}}_{\mathcal{H}(\mathbf{r}, \mathbf{p}, D; \psi, E^*)} \\
    &\equiv \inf_D H(\mathbf{r}, \mathbf{p}, D; E^*),
    \label{eq:legendre}
\end{split}
\end{equation}
where $H_0$ is the microscopic Hamiltonian depending on the positions $\mathbf{r}$ and momenta $\mathbf{p}$ of all atoms and the electronic state $\psi$, $\Omega = AL$ is the unit-cell volume, $\varepsilon_0$ is the permittivity of the vacuum, and $P(\mathbf{r; \psi})$ is the total polarization of the system at given $\mathbf{r}$.
$\psi$ is a representation of electronic variables and may denote, for example, the electronic wavefunction, the electronic density or a coarse-grained representation such as the \ac{cpm} charges.
This electronic dependence is used in the derivation in Appendix~\ref{app:theory-eom}, but it is not necessary to make it explicit for the discussion regarding the force-field implementation of the electrostat discussed below.
We thus work with the Legendre objective function $H$ as a quantity already minimized with respect to $\psi$. We note that the Legendre term $\Omega DE^*$ in \cref{eq:legendre} quantifies the thermodynamic work of the voltage source needed to transfer an amount of charge between the opposing cell boundaries (\textit{cf}. the circuit scheme in \cref{fig:cell-definition}A).

The infimum in~\cref{eq:legendre} is attained at~\cite{griffiths_introduction_1989} 
\begin{equation}
    D = \varepsilon_0E^* + P,
    \label{eq:def-D}
\end{equation}
thus resulting in
\begin{equation}
\begin{split}
H_E\left(\mathbf{r}, \mathbf{p}; E^*\right) 
    &=
    H_0\left(\mathbf{r}, \mathbf{p}\right) 
    -
    \frac{\Omega\varepsilon_0}{2}{E^*}^2 \\
    &- 
    \Omega P(\mathbf{r}) E^*,
    \label{eq:ssv}
\end{split}
\end{equation}
where $H_0$ and $P$ are again understood as evaluated at the minimum with respect to $\psi$.
In \cref{eq:ssv}, the quadratic term in $E^{*}$ represents the negative of the electrostatic self-energy of the field per unit cell and adopts, naturally, a constant value under constant-$E$ conditions. It is normally neglected as a trivial energy offset in constant-volume \ac{md}~\cite{umari_ab_2002}.

A strictly constant $E^*$ measured across the cell is valid only in the thermodynamic limit, and a finite-size simulation must accommodate thermal fluctuations of the control variable.
Clearly, $H_E(\mathbf{r}, \mathbf{p}; E^*)$ cannot describe such ventures. 
However, $H(\mathbf{r}, \mathbf{p}, D; E^*)$ (defined in \cref{eq:legendre}) is the proper function to describe the energy of the system when $E \equiv \varepsilon_0^{-1}(D - P) \neq E^*$.
One can show that 
\begin{equation}
    \frac{\partial H}{\partial D}
    =
    \Omega (E - E^*)
\end{equation}
is the appropriate form of the generalized force that drives the system towards $E^*$ (see Appendix~\ref{app:theory-eom}).
Defining and controlling the magnitude of the variation of $D$ (and consequently of $E$), requires an equilibrium thermal phase-space probability distribution at temperature $T$
\begin{equation}
   \rho(x; E^*) \propto e^{-\beta H(x; E^*)}
   \label{eq:ensemble}
\end{equation}
with $x \equiv (\mathbf{r}, \mathbf{p}, D)$ and $\beta = (k_\mathrm{B}T)^{-1}$.

To derive a practical propagation procedure for $E$ under this distribution, we rely on the correspondence between the diffusive Fokker--Planck dynamics of ensemble-densities and stochastic phase-space dynamics described by It\^{o} \acp{sde}~\cite{gardiner_hanbook_1985, kloeden_numerical_1999}.
We present a detailed derivation in Appendix~\ref{app:theory-eom} and additional details in Section~\ref{SI-sec:es-derivation} of the SI.

The relevant It\^{o} \ac{sde} reads
\begin{equation}
\begin{split}
    \dd E
    &=
    -\frac{1}{\tau_0} \left(E - E^*\right) \ \dd t \\
    &- \varepsilon_0^{-1} j \ \dd t \\
    & + \sqrt{\frac{2}{\tau_0}\frac{k_\mathrm{B}T}{\varepsilon_0 \Omega}} \ \dd W_t, 
    \label{eq:ou-process}
\end{split}
\end{equation}
where $j = \dd P / \dd t$ is the current density, $\tau_0$ is the time constant, and $W_t$ represents a Wiener process~\cite{gardiner_hanbook_1985}.
The first and last terms of \cref{eq:ou-process} describe the dissipation and fluctuation of $E$, respectively.
The scale of the fluctuation follows from the stationarity condition of $\rho(x, E^*)$ and does not have to be known \textit{a priori}. 
The term involving the current $j$ explicitly describes the polarization response of the system and couples the motion of the system to the dynamics of $E$.

\cref{eq:ou-process} is fully general and makes no assumptions about the composition of the system or the nature its polarization response.
However, in order to obtain a simple practical scheme in the context of empirical force fields,
we start by assuming linear polarization with an electronic susceptibility $\chi$
\begin{equation}
    P(\mathbf{r})
    =
    P(\mathbf{r})\Big|_{E=0}
    + \varepsilon_0\chi E
    =
    \frac{P_\mathcal{E}(\mathbf{r})}{g}
    +
    \varepsilon_0 \left(\frac{1}{g} - 1\right) E,
    \label{eq:linear-polarization}
\end{equation}
where the second equality, derived in Appendix~\ref{app:theory-es-vs-ps}, assumes an electrolyte polarization $P_\mathcal{E}(\mathbf r)$ described by a non-polarizable force-field, and an ideal metallic behavior of the model planar electrode.
In the last expression, 
\begin{equation}
    g = \frac{L'}{L} \label{eq:geo-term}
\end{equation}
is a geometric property of the system and $L'$ is the capacitor length (see \cref{fig:cell-definition}B).
Expressing $j$ through \cref{eq:linear-polarization} directly yields
\begin{equation}
\begin{split}
    \dd E
    &=
    -\frac{1}{\tau_0'} \left(E - E^*\right) \ \dd t \\
    &- \varepsilon_0^{-1} j_\mathcal{E} \ \dd t \\
    & + \frac{1}{L}\sqrt{\frac{2}{\tau_0'}\frac{k_\mathrm{B}T}{C_0}} \ \dd W_t, 
    \label{eq:ou-process-deissenbeck}
\end{split}    
\end{equation}
where $j_{\mathcal{E}}$ is the electrolyte current, the constant $\tau_0$ has been scaled according to  $\tau_0' = \tau_0/g$ and $C_0 = \varepsilon_0A/L'$ represents the vacuum capacitance of the system.

The specific form of the SDE represents a driven Ornstein--Uhlenbeck process and can be solved exactly by It\^{o}'s lemma~\cite{oksendal_stochastic_1998} with a finite time step $\Delta t$ (see derivation details in Section~\ref{SI-sec:ito-solution} of the SI), yielding
\begin{equation}
\begin{split}
    E(t + \Delta t)
    &=
    E^* + \alpha \left[E(t) - E^* \right] \\
    & - \tau_0'\left( 1 - \alpha \right) \varepsilon_0^{-1} j_{\mathcal{E}}(t) \\
    & + \mathcal{N}(t)\frac{1}{L}\sqrt{\frac{k_\mathrm{B}T}{C_0}\left( 1 - \alpha^2 \right)}    
    \label{eq:es-eom}
\end{split}
\end{equation}
where we made the time dependence of different variables explicit, $\alpha = e^{-\Delta t / \tau_0'}$ and $\mathcal{N}$ is a random Gaussian number with zero mean and unit variance.
This is the working equation of the electrostat in this work, describing the stochastic dynamics of $E$ that ensure the required ensemble mean $\langle E \rangle = E^*$ with consistent thermodynamic fluctuations.

\cref{{eq:es-eom}} represents a finite-field counterpart of the original stochastic potentiostat~\cite{deisenbeck_dielectric_2021}.
The two methods formally coincide in the $\Delta t \rightarrow 0$ limit: the original approach treats the current term implicitly and through a finite difference, approximating the $\tau_0'(1 - \alpha)$ prefactor with $\Delta t$. 
This correspondence is derived in Appendix~\ref{app:theory-es-vs-ps} and the resulting deviation between the two methods is elaborated in Section~\ref{SI-sec:eq-es-tps} of the SI.

\subsection{Propagating an electrostatted \ac{md} simulation}
\label{sec:propagator}

The electrostat introduces the $E$-field as a stochastic dynamical variable with an evolution given by \cref{eq:es-eom}.
In order to generate a practical finite-time-step propagator, we relied on the discretized Liouville operator $\mathcal{L}$ formalism~\cite{bussi_accurate_2007} to generate a trajectory of the state point $x(t)$.
In the present case,
\begin{equation}
    \mathcal{L}
    =
    \mathcal{L}_\mathrm{A}
    +
    \mathcal{L}_\mathrm{B}
    +
    \mathcal{L}_\mathrm{ES}
    =
    \bm{\nabla}_\mathbf{p}H \cdot \bm{\nabla}_\mathbf{r} 
    - 
    \bm{\nabla}_\mathbf{r}H \cdot \bm{\nabla}_\mathbf{p}
    +
    \mathcal{L}_\mathrm{ES},
\end{equation}
where $H$ is given by \cref{eq:ssv}, $\mathcal{L}_\mathrm{A} = \dot{\mathbf{r}} \cdot \bm{\nabla}_\mathbf{r}$ and $\mathcal{L}_\mathrm{B} = \mathbf{F} \cdot \bm{\nabla}_\mathbf{p}$ are the deterministic position and momentum terms, $\mathcal{L}_\mathrm{ES}$ is responsible for the stochastic propagation of the $E$-field and, finally, $\mathbf{F} = -\bm{\nabla}_\mathbf{r}H(x; E^*)$ are the atomic forces.
Note that the identical forces can be obtained by directly differentiating the constant-$E$ Hamiltonian (last line of \cref{eq:ssv}) with $E^* = E$, which is the practical choice for the implementation (see Appendix~\ref{app:theory-eom}).

Different propagators can be obtained systematically by different Trotter splitting choices of the exponential propagator $e^{\mathcal{L}(t - t_0)} \approx \left(e^{\mathcal{L}\Delta t}\right)^P$ as $P$ gets large.~\cite{tuckerman_statistical_2010} The original potentiostat implementation~\cite{deisenbeck_dielectric_2021} as well as its implementation in \textsc{Lammps/Electrode}~\cite{ahrens-iwers_electrode_2022} employs an approach that can be smoothly integrated into the existing infrastructure without the need to modify its MD propagators. 
Specifically, this is a symmetric split of the B-term combined with an asymmetric split of the ES- and A-terms, which commute
\begin{equation}
    e^{\mathcal{L}\Delta t}
    \approx
    e^{\mathcal{L}_\mathrm{B}\frac{\Delta t}{2}} e^{\mathcal{L}_\mathrm{ES}\Delta t} e^{\mathcal{L}_\mathrm{A}\Delta t} e^{\mathcal{L}_\mathrm{B}\frac{\Delta t}{2}}.
\end{equation}
When applied to an initial condition $x(t)$,  the following explicit sequence of update steps is generated for a time step $\Delta t$:
\begin{equation}
\begin{aligned}
\text{(i)}\quad&
\mathbf{p}\left(t+\frac{\Delta t}{2}\right)
=
\mathbf{p}(t)
+\frac{\Delta t}{2}\mathbf{F}(t),
\\[0.4em]
\text{(ii)}\quad&
\mathbf{r}(t+\Delta t)
=
\mathbf{r}(t)
+\Delta t\,
\dot{\mathbf{r}}\left(t+\frac{\Delta t}{2}\right),
\\[0.4em]
\text{(iii)}\quad&
E(t+\Delta t)
\;\text{from \cref{eq:es-eom}},
\\[0.4em]
\text{(iv)}\quad&
\mathbf{F}(t+\Delta t)
\;\text{from }E(t+\Delta t)
\text{ and }\mathbf{r}(t+\Delta t),
\\[0.4em]
\text{(v)}\quad&
\mathbf{p}(t+\Delta t)
=
\mathbf{p}\left(t+\frac{\Delta t}{2}\right)
+\frac{\Delta t}{2}\mathbf{F}(t+\Delta t).
\end{aligned}
\end{equation}
This corresponds to a B-ES-A-B propagator structure equivalent to a velocity-Verlet propagator with the $E$-field update placed after the position update.
We note that the reason that a symmetric split of the ES-term here would not change the dynamics in any way is that the electrostat update operator $\mathcal{L}_\mathrm{ES}$ commutes with the position update operator $\mathcal{L}_\mathrm{A}$.
This property is demonstrated in Section~\ref{SI-sec:symmetric-propagator} of the SI.

As suggested by the required ensemble distribution in \cref{eq:ensemble}, the electrostat defined above is able to enforce a constant temperature of a simulation on its own, as demonstrated in Section~\ref{SI-sec:validation} of the SI.
However, it can be additionally seamlessly integrated to a thermostatted simulation for a more tailored and robust temperature control~\cite{deisenbeck_dielectric_2021}.

\subsection{Implementation in LAMMPS}

We build the \textsc{Lammps/Electrode} implementation of the electrostat based on the existing \ac{cpm} and constant-$E$ infrastructure~\cite{ahrens-iwers_electrode_2022}, which we briefly summarize in the following paragraph.
The \ac{cpm} models the ideal conductive electrodes as a set of fluctuating Gaussian charges $\mathbf q \in \mathbb{R}^N$ centered on the $N$ electrode atoms under 2D-\ac{pbc}.
The energy of that arrangement is
\begin{equation}
    U_\mathrm{CPM}(\mathbf q) = \frac12 \mathbf q^\mathrm T \mathbf A \mathbf q + \mathbf b^\mathrm T \mathbf q - \mathbf v^\mathrm T \mathbf q
    \label{eqn:Uele}
\end{equation}
where the $\mathbf A$-matrix contains the Coulomb interactions between electrode atoms, the vector $\mathbf b$ contains the corresponding electrode-electrolyte interactions and the vector $\mathbf{v}$ serves to enforce the Dirichlet boundary condition and contains constant entries for each of the two electrodes individually, defining the potential drop $\phi^*$ between them.
The optimal charges $\mathbf{q}'$, consistent with the constant-potential condition, are obtained through minimizing the energy
\begin{equation}
    \frac{\partial U_\mathrm{CPM}}{\partial q_i}
    =
    0
    \implies 
    \mathbf{q}'
    =
    \mathbf{A}^{-1}\left( \mathbf{v} - \mathbf{b} \right)
\end{equation}
for each geometry.
The charge neutrality of the entire system can be enforced through adding an offset $v_0$ to the potential defined in $\mathbf{v}$ as a Lagrange multiplier~\cite{nakano_chemical_2019}.

The constant-$E$ method is obtained from the \ac{cpm} formulation by introducing full \acp{pbc} and accordingly modifying the energy of the charge arrangement to reflect the presence of the homogeneous field $E^*$~\cite{dufils_simulating_2019, ahrens-iwers_electrode_2022}
\begin{equation}
    U(\mathbf{q})
    =
    U_\mathrm{CPM}(\mathbf{q}) -  \mathbf{q}^T\mathbf{z} E^*.
    \label{eq:Uce}
\end{equation}
In this case, the use of full periodicity merges the two electrodes into a single one and $\mathbf{v}$ takes on identical entry $v_0$ for all the electrode atoms.
Minimizing \cref{eq:Uce} gives the following expression for the optimal set of charges
\begin{equation}
    \mathbf{q}'
    =
    \mathbf{A}^{-1}\left( \mathbf{v} + \mathbf{z}E^* - \mathbf{b} \right).
\end{equation}
The technical details of the constant-$E$ implementation in \textsc{Lammps/Electrode} can be found in Section~\ref{SI-sec:electrode} of the SI.

\subsubsection*{Constant-$D$ simulations}

As a new addition to the \textsc{Electrode} package we have also implemented the constant-$D$ method. 
See discussion in Appendix \ref{app:theory-eom}.
The implementation follows the procedure of applying an appropriate $E$-field which varies in time, given by
\begin{equation}
    E(t)
    =
    \varepsilon_0^{-1}\left[D^* - P(t)\right]
    =
    \varepsilon_0^{-1} \left[gD^* - P_\mathcal{E}(t)\right],
    \label{eqn:eff-D}
\end{equation}
which is imposed at every time step through the constant-$E$ approach (see Appendix~\ref{app:theory-equivalence} for details).
The second equality was obtained by applying \cref{eq:linear-polarization}.
The electrolyte polarization
\begin{equation}
    P_\mathcal{E}(t)
    =
    \Omega^{-1}\sum_{i \in \mathcal{E}} q_i z_i(t)
\end{equation}
is calculated simply at each time step via an itinerant sum~\cite{caillol_comments_1994} over all atoms belonging to the electrolyte $\mathcal{E}$.
Additional options of calculating $P_\mathcal{E}$ are discussed in Section~\ref{SI-sec:dfield} of the SI.
The constant-$D$ method coincides with \ac{ccm} when one chooses $D^* = Q^* / A$.

\subsubsection*{Electrostat}

Finally, we have implemented the electrostat through \cref{eq:es-eom} in the \textsc{Electrode} package also exploiting the existing constant-$E$ infrastructure with a variable field magnitude.
The required electrolyte current is calculated in the context of a fixed-charge force field as a time derivative of the polarization
\begin{equation}
    j_\mathcal{E}(t)
    =
    \frac{\dd P_\mathcal{E}(t)}{\dd t}
    =
    \Omega^{-1}\sum_{i \in \mathcal{E}} q_i \dot{z}_{i}(t),
\end{equation}
which depends only on atomic velocities.
The equation of motion \cref{eq:es-eom} relies on the prior knowledge of the vacuum capacitance $C_0$ that parametrizes the stochastic term.
In our implementation, this is obtained fully consistently from the electrode interaction matrix $\mathbf{A}$ as described in Section~\ref{SI-sec:implementation-efield} of the SI.
As such, executing an \ac{md} simulation with the electrostat requires no additional parameters except for the temperature and the time constant $\tau_0'$ as inputs.

In spite of its technical feasibility, we do not implement an electrostat based on \cref{eq:ou-process} in this work.
The reason is that the linear polarization and ideal metallic electrode assumptions are exact for the \ac{cpm} formalism. 
As convincing numerical evidence, we demonstrate in Section~\ref{SI-sec:additional-results} of the SI that the total current $j$ along an electrostatted simulation can be reconstructed exactly by combining $j_\mathcal{E}$ and $E(t)$, confirming \cref{eq:linear-polarization}.
An implementation based on \cref{eq:ou-process} is relevant when going beyond non-polarizible empirical force fields, as elaborated in Section~\ref{sec:Conclusion}.

The electrostat can be coupled to an explicit thermostat.
Specifically, the native canonical \ac{md} infrastructure of \textsc{Lammps} implements a Martyna--Klein--Tuckermann O-B-A-B-O propagator~\cite{martyna_nosehoover_1992} for Nos\'{e}--Hoover chains~\cite{nose_unified_1984, hoover_canonical_1985} (where O represents the update of the thermostat chain): here, the electrostat can be easily inserted as an O-B-ES-A-B-O sequence, consistently with the microcanonical case described above.
The compatibility with other thermostats (such as local or global stochastic ones) is implied~\cite{deisenbeck_dielectric_2021}.

\section{\label{sec:Results and Discussion} Results and Discussion}

\subsection{Ensemble equivalences}
\label{sec:benchmarks}

\begin{figure}[tb!]
    \includegraphics{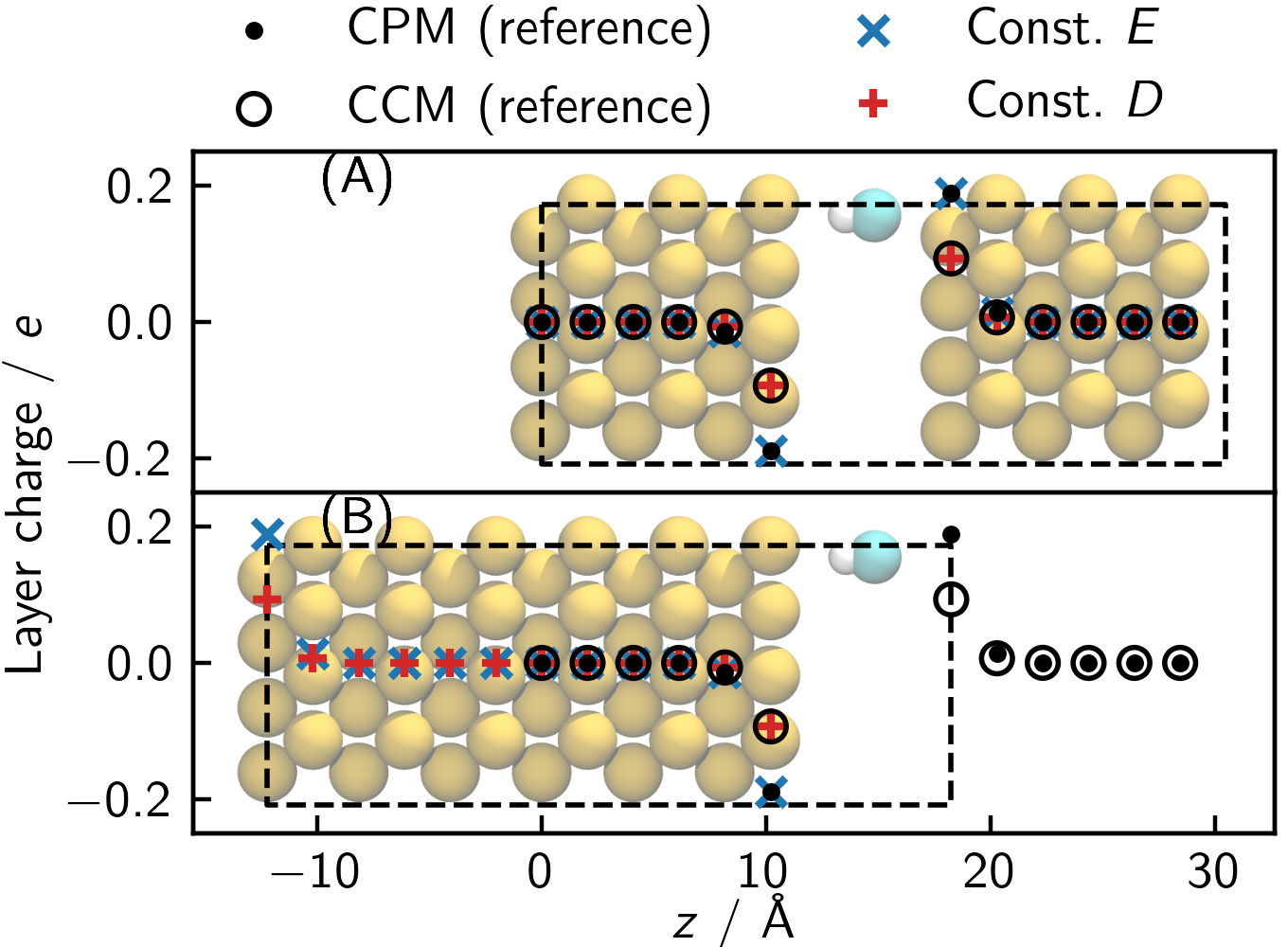}
    \caption{
        Equivalences of different electrostatic descriptions of a model HCl/Au(100) nanocapacitor.
        In panel A, we show the electrolyte-centered system and the charge per layer as calculated by the four methods discussed above in the text.
        The data shows that the \ac{cpm} and the constant-$E$ methods as well as the \ac{ccm} and the constant-$D$ methods represent equivalent descriptions under their respective \acp{pbc}.
        In panel B, we show the same system wrapped into the conductor-centered geometry.
        In this case, the \ac{cpm} and \ac{ccm} are not applicable and both sets of  black-point references are reproduced from panel A.
        For the constant-$E$ and -$D$ methods, the data shows that the finite-field description is invariant to translation within the unit cell and correctly gives charges consistent with the reference.
        A snapshot of the system geometry is shown in each panel, with Au atoms in gold, Cl in turquoise and H in white.
        The unit-cell dimensions are indicated by the black dashed lines.
    }
    \label{fig:layer-charges}
\end{figure}
In the following Section, we present numerical benchmarks of the equivalences of the various electrochemical ensembles introduced in \cref{fig:cell-definition} using our new implementation in \textsc{Lammps/Electrode}.

First, we benchmark the per-layer charge in a simple model system consisting of an Au(100) slab nanocapacitor with a single hydrogen chloride molecule as the dielectric filling in various system configurations obtained as a single-point evaluation.
These results are summarized in \cref{fig:layer-charges}.
In \cref{fig:layer-charges}A, we show the electrolyte-centered geometry containing two individual electrodes each consisting of 6 Au layers.
The \ac{cpm} method with $\phi^* = \SI{2}{\volt}$, which treats the system as open in the $z$ direction, deposits a charge of approximately $\pm0.2$~e at the interface, which abruptly decays into the bulk as expected for a conductor.
In the constant-$E$ method the system is fully periodic (the unit cell is shown as the black dashed line), and the $E$-field was set to a magnitude of \SI{65.6}{\milli\volt \angstrom^{-1}} that reproduces the potential of \SI{2}{\volt} across the unit-cell length.
The \ac{ccm} was set to target a total electrode charge of $Q = \SI{0.1}{\elementarycharge}$ and, correspondingly, the displacement field in the constant-$D$ method was set to \SI{0.205}{\milli\elementarycharge{} \angstrom^{-2}} to obtain the same charge.
The closed- and open-circuit ensembles provide different values of the per-layer charge, but remain internally consistent: \ac{cpm} is equivalent to constant-$E$ and \ac{ccm} to constant-$D$ as expected. 

In \cref{fig:layer-charges}B, we show the same system but translated in the periodic unit cell by half a cell-length in the $z$-direction.
In this configuration, the \ac{cpm} and \ac{ccm} are no longer applicable, but the field-based constant-$E$ and constant-$D$ methods yield per-layer charges identical as above, since translating the periodic system within the unit cell does not change any physics.
Adding a vacuum layer within the electrode is technically possible with our implementation and is discussed in \cref{SI-fig:benchmark-vacuum} of the SI.

\begin{figure}[tb!]
    \centering
    \includegraphics[width=\linewidth]{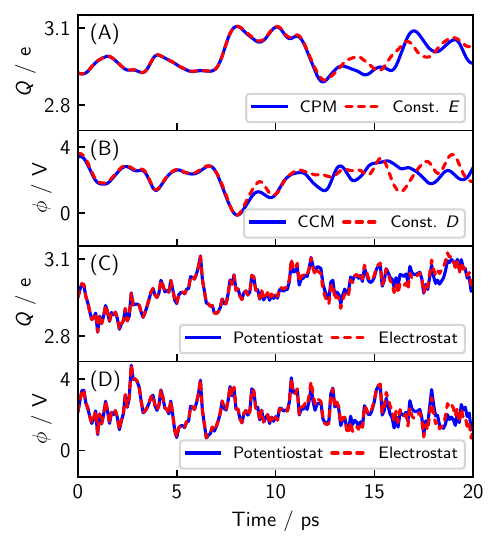}
    \caption{
    Equivalence of the 2D- vs 3D-\ac{pbc} simulations in the [BMIM$^+$][PF$^-_6$]/graphite system (\cref{fig:cell-definition}).
    Panel A compares the evolution of total electrode charge in a constant-potential simulation using the \ac{cpm} (blue) and the constant-$E$ method (red); both simulations started from the same initial condition and the voltage drop is \SI{2}{\volt}, which was enforced directly in \ac{cpm} or by an $E$-field of the corresponding magnitude (detailed in the text).
    Panel B shows a similar comparison between \ac{ccm} with the charge constrained to \SI{3}{\elementarycharge} and the constant-$D$ method with a correspondingly strong $D$-field (see text).
    Panel C compares the charge dynamics in stochastic simulations where the voltage drop of 2~V is maintained by the potentiostat or electrostat.
    Panel D shows the evolution of the potential in the same simulation as shown in panel C. 
    In all red curves, the values of the monitored quantities were calculated from the corresponding magnitudes of the finite fields using relations discussed above and in Appendix~\ref{app:theory-equivalence}.
    }
    \label{fig:equivalence}
\end{figure}

The results above are obtained for single-point calculations. 
However, the discussion in Appendix~\ref{app:theory-equivalence} suggests an even stronger statement: the electrostatic potential between the electrodes in the \ac{cpm} and constant-$E$ case is the same for every electrolyte configuration, therefore the dynamics generated under the respective approaches should also be identical.
We numerically verify this statement  on the ionic liquid supercapacitor shown in \cref{fig:cell-definition} and summarize the findings in \cref{fig:equivalence}.

In \cref{fig:equivalence}A, we compare the evolution of the electrode charge in two simulations started from exactly the same initial condition and evolved using the \ac{cpm} or constant-$E$ methods.
For the former, we employed a potential of \SI{2}{\volt}, for the latter an electric field of \SI{15.6}{\milli\volt\per\angstrom}, which corresponds to a \SI{2}{\volt} potential for the system in question with $L = $ \SI{128.3}{\angstrom}.
For the first approximately \SI{12}{\pico\second} ($\sim$6000 steps, see Section~\ref{SI-sec:computational} of the SI for details), the two trajectories evolve  identically, after which they start to diverge.
The initial agreement is a manifestation of the same electrostatics in the two cases that lead to the same forces in each configuration; the divergence is explained by the onset of a Lyapunov instability~\cite{frenkel_understanding_2002, tee_fully_2022}: an exponential build up of error due to the minute numerical differences in the respective propagators.
In \cref{fig:equivalence}B, we show the same for the fluctuating potential with the electrode charge constrained to \SI{3}{\elementarycharge} via \ac{ccm} and using a displacement field of \SI{2.39}{\milli\elementarycharge\per\angstrom\squared} for the constant-$D$ approach, which was set to reproduce \SI{3}{\elementarycharge}. 
In \cref{fig:equivalence}C and \ref{fig:equivalence}D we compare in a similar way the potentiostat and the  electrostat (again, employing a target $E^*$ with an appropriate magnitude); here we show both $Q$ and $\phi$, respectively, as both quantities fluctuate in the stochastic trajectory (the sequence of the random number generator was set to be the same in both cases).
Again, the same trend is obtained where the simulations perfectly overlap over many integration time steps before diverging, as the numerical instability sets in.

These results offer solid numerical evidence that the \ac{cpm} and \ac{ccm} may be used interchangeably with constant-$E$ and constant-$D$ methods, as well as the potentiostat with electrostat.
In the rest of this paper, we rely only on the 3D-periodic simulations with a fixed, or a fluctuating homogeneous field.

\subsection{Ion-separation free-energy profiles at electrochemical interfaces}

\begin{figure}[tb!]
    \includegraphics[width=\linewidth]{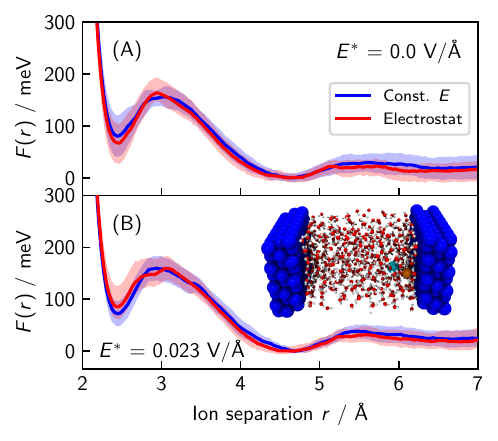}
    \caption{
    Helmholtz free-energy $F$ profiles as a function of the separation $r$ of a LiCl ion pair at the water/Pd(111) interface.
    Panel A: The free-energy profile obtained by the constant-$E$ method at grounded (zero-field) conditions is shown in blue, the corresponding one obtained by the electrostat is shown in red.
    The error bars are plotted at shaded intervals behind each profile in matching color.
    Panel B: The same calculation as in panel A, only with a \SI{0.023}{\volt\per\angstrom} target field applied.
    The inset shows a snapshot of the system in which the Pd electrodes are shown in blue, oxygen atoms in red, hydrogen atoms in white, the Li$^+$ ion in orange and the Cl$^-$ ion in turquoise.
    The system is fully periodic, identically to \cref{fig:cell-definition}B.
    }
    \label{fig:ions}
\end{figure}

As an exemplary application of the methodology, we simulated the free-energy profiles of the separation of a LiCl ion pair where the cation is held adsorbed at the interface of water and a model Pd electrode under an electric field (see snapshot in \cref{fig:ions}B) to model a voltage bias. 
The field was imposed using the electrostat, and compared to the constant-$E$ method without field fluctuations.
The free-energy profiles were calculated using the umbrella sampling method~\cite{torrie_nonphysical_1977, kumar_weighted_1992} with a total of 200~ps of simulated time per window.
The complete description of the simulations is presented in Section~\ref{SI-sec:computational} of the SI and the main results are shown in \cref{fig:ions}.
The two panels compare the free-energy profiles at zero field (grounded electrodes) and a field of \SI{0.023}{\volt\per\angstrom} which corresponds to a voltage drop of \SI{1}{\volt} for this system. 
In all cases, the free-energy profile exhibits a structure typical for the separation of oppositely charged ions in an aqueous environment consisting of two minima and a barrier between them.
The first, higher-lying minimum of the free-energy profile corresponds to the so-called \ac{cip}, where the ions directly touch; the second lower-lying minimum corresponds to the \ac{ssp}, in which a water molecule is located between the two ions.
The height of the central barrier relative to the \ac{ssp} minimum is roughly 160~meV (\ie{}, 15.4~kJ/mol), a value that quantitatively corresponds to existing studies of ion association with the same class of water force fields~\cite{zhang_lithium_2004, kohagen_exploring_2015}.
From a physical point of view, we do not observe a difference in ion association between the charged and grounded interface.
This outcome can be attributed to the expected screening of the electrode charge by the \ac{edl}, which makes the bulk regions of in two simulations similar from an electrostatics perspective.

Turning our attention to the methodological comparison of the electrostat and constant-$E$ curves, we note that within the calculated error bars, the two free-energy profiles in each panel of \cref{fig:ions} cannot be considered distinct: the slight differences between the profiles are of statistical origin.
In Section~\ref{SI-sec:additional-results} of the SI, we demonstrate that the corresponding simulation obtained under 2D-\ac{pbc} using the \ac{cpm} and the potentiostat leads to the same conclusion.
This implies that the electrolyte effectively experiences the same free-energy landscape irrespective of the presence (or absence) of noise in the external field, and, consequently, samples the same ensemble of structures.
These results motivate acquiring a deeper understanding of the conditions that require including the correct fluctuational physics through the electrostat and those that do not.
In the rest of this paper, we will further illustrate this question using a different example and analyze it in more detail.

\subsection{Diffusion in [BMIM$^+$][PF$^-_6$]/graphite under applied voltage}
\label{sec:diffusion}

\begin{figure}
    \includegraphics[width=\linewidth]{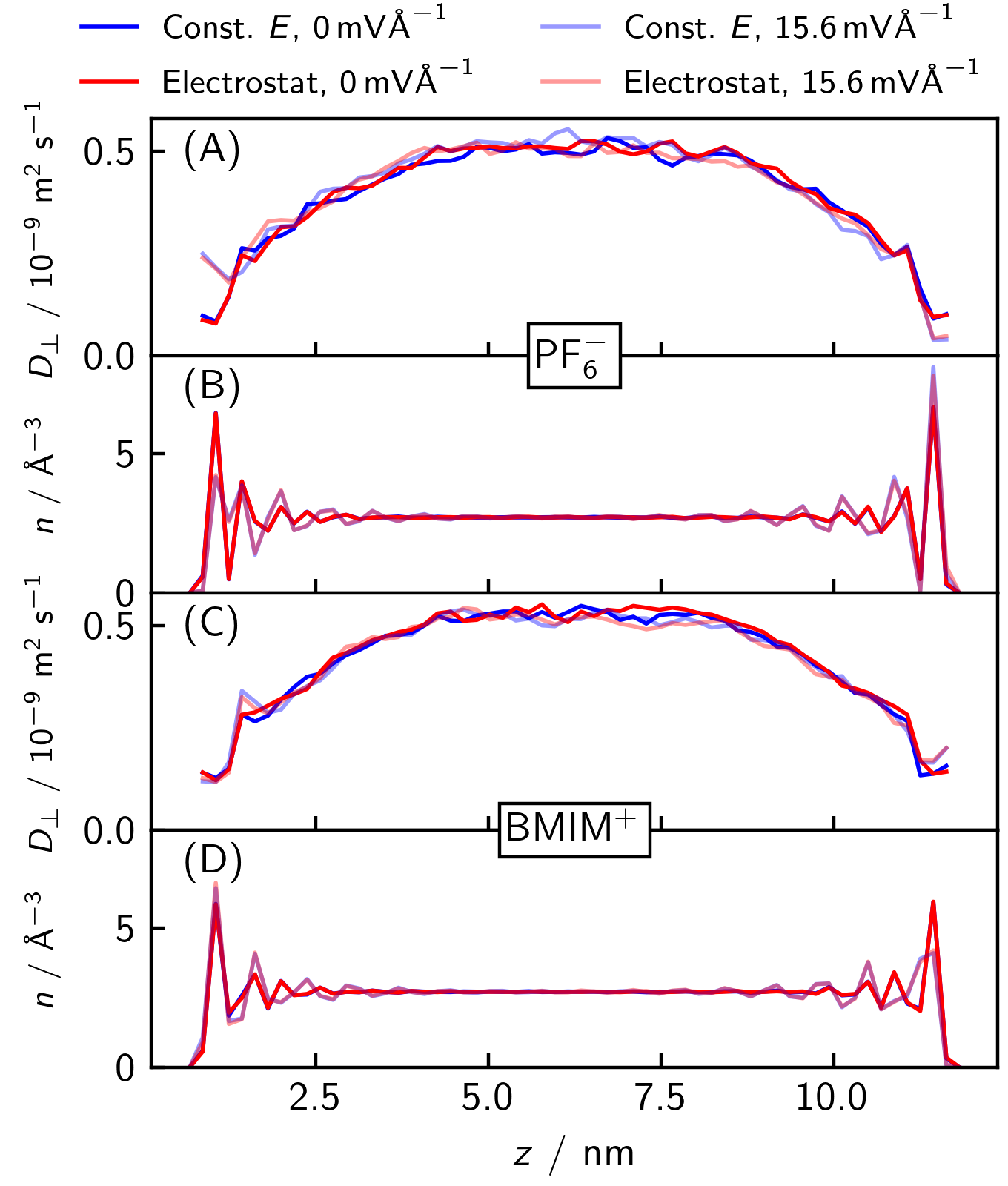}
    \caption{
    Diffusion coefficients $D_\perp$ for the motion in $z$-direction as a function of $z$ for \ce{PF6-} (A) and \ce{BMIM+} (C) ions in a model system with an ionic liquid and graphite electrodes (\cf{} \cref{fig:cell-definition}).
    Number-density profiles are shown in (B) and (D) for \ce{PF6-} and \ce{BMIM+}, respectively. 
    The coefficients and densities are shown for $E$-field of \num{0} and \SI{15.6}{\milli\volt\per\angstrom}, which were realized with constant-$E$ and the electrostat.
    }
    \label{fig:diffusion}
\end{figure}

At this point, we return to the ionic liquid supercapacitor system shown in \cref{fig:cell-definition}.
Specifically, our interest lies in describing the diffusion of the constituent ions in the normal ($z$) direction at the electrode-electrolyte interface and answering whether the fluctuations imposed by the electrostat affect the mobility near the interface in comparison to constant-$E$ conditions and manifest as changes in diffusion coefficients.
These were calculated using the Einstein diffusion formula (see details in Section~\ref{SI-sec:computational} of the SI) evaluated on \ac{md} trajectories spanning the total of \SI{200}{ns} of simulated time and performed at \num{0} and \SI{15.6}{\milli\volt\per\angstrom} fields (the latter corresponding to a \SI{2}{\volt} voltage), as enforced by both control approaches.
A detailed description of the performed simulations can be found in Section~\ref{SI-sec:computational} of the SI.

The calculated normal-direction-resolved diffusion coefficients $D_\perp(z)$ and the corresponding number-density profiles along $z$ are shown in \cref{fig:diffusion}.
The spatially resolved profile of the diffusion coefficients is characterized by a concave shape (\cref{fig:diffusion}A and C for the anion and cation, respectively): they assume small values at the interface and grow to a plateau in the bulk region between the electrodes.
This plateau corresponds to the flat middle region of the density profiles (\cref{fig:diffusion}B and D) in which the ionic liquid displays a bulk character and, conversely, the decrease of $D_\perp$ towards the electrodes is caused by the structuring of the interfacial liquid. 
The values near the interface differ between the two fields. 
This effect is the most notable for PF$_6^-$, which has a higher charge density: at the opposite charged anode (right), the species is strongly adsorbed and slowed down, as opposed to the like-charged interface (left) from which it is repelled.
At zero bias, the spatial dependence of $D_\perp$ is symmetric.
This behavior is directly reflected in the density profiles.
At \SI{15.6}{\milli\volt\per\angstrom}, the anion density in panel \cref{fig:diffusion}B is increased at the anode and decreased at the cathode; for the cations the effect is reversed.
At zero field, the profiles remain symmetric, too.

It is worth noting that the calculated spatial profiles are once again the same, within statistical uncertainty,  between the electrostat and constant-$E$ method.
Our results thus show that interfacial diffusion, like the free-energy profiles discussed in the previous Section, remains agnostic to the fluctuations of the field at the simulated system size.

\subsection{When is it important to use an electrostat?}
\label{sec:size-effect}

In the light of the findings of the two preceding Sections, the last point in this discussion that remains to be answered is the threshold of the system size at which using the electrostat actually starts to make a difference in comparison to constant-$E$ simulations.
To address this, we probe the electrode charge $Q$, which is tied to the core quantities of \cref{eqn:Uele} and affects the dynamics directly.
We note that the standard deviation of the electrode charge density scales with the surface area of the electrode $A$ (\textit{cf.} its definition in \cref{fig:cell-definition}D) as $\propto \sqrt{C_0}/A \ \propto 1 / \sqrt{A}$ at fixed $L'$: this is consistent with the fact that thermodynamic fluctuations vanish in the thermodynamic limit.
We do not address the dependence on $L'$ here, which is known to be very weak owing to \ac{edl} formation~\cite{guedes_does_2026}.
Effectively, by studying a series of representations of the system in question at systematically changing $A$, we can estimate the practical onset of where the electrostat and constant-$E$ descriptions start to differ.

\begin{figure}[tb!]
    \includegraphics[width=\linewidth]{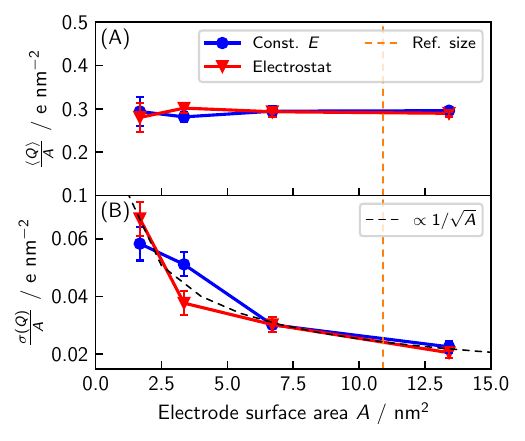}
    \caption{
    Scaling of charge fluctuation with the electrode surface area.
    Panel A: the curves explicitly demonstrate the lack of dependence of the mean charge density ($\langle Q \rangle / A$) on $A$ for constant-$E$ (blue) and the electrostat (red).
    The error bars represent the standard error of the mean obtained by block averaging and show at a $3\sigma$-confidence interval.
    The orange vertical line shows the area of the reference system shown in \cref{fig:cell-definition} and studied in Section~\ref{sec:diffusion}.
    Panel B shows the dependence of the standard deviation of the electrode charge density $\sigma(Q) / A$ as a function of $A$.
    Same color coding applies and the error bars have the same meaning as in the panel above.
    The dashed black line shows a curve with a $1/ \sqrt{A}$ functional form fitted to the electrostat data points. 
    }
    \label{fig:fluctuations}
\end{figure}

Such a calculation is shown in \cref{fig:fluctuations} for four size representations of the ionic liquid supercapacitor held under a field of \SI{15.6}{\milli\volt\per\angstrom}, and with a $A$ ranging from roughly 1.7 to \SI{13.4}{\nano\meter\squared} at constant $L$~\cite{blundell_concepts_2017}.
In \cref{fig:fluctuations}A, we show that the mean charge density $\langle Q \rangle/A$ stays constant with increasing $A$, which is equally true for both approaches.
In \cref{fig:fluctuations}B, we present the dependence of the standard deviation of the electrode charge density $\sigma(Q)/A$ on $A$.
The electrostat data follows the $1/\sqrt{A}$ fit with a good agreement over the studied range.
The constant-$E$ data agree with the electrostat data for the two largest system sizes, and the size of the system discussed in Section~\ref{sec:diffusion} (marked by the orange line) falls well into this regime.
As the electrode area gets smaller, the curves start to deviate from each other.
At first, this can be attributed to the fact that the charge distribution, dominated at large system sizes by the fluctuation of $P$, begins to feel a stronger contribution from the stochastic fluctuations.
This roughly corresponds to the region between $\sim3$ and \SI{7}{\nano\meter\squared}, where  constant-$E$ is indeed an incomplete description.
The behavior at even smaller system sizes is likely burdened by a lateral finite-size effect, where the liquid character and mobility of the electrolyte are compromised (note that the smallest system is only about $\num{9}\times\SI{17}{\angstrom}^2$ in area).
At that stage, the system stops being physical and the fluctuations under constant-$E$ start to flatten out; the electrostat provides strong stochastic fluctuations from the device by construction, but the dynamics of the system remain unphysical. To provide further insight into the matter, in Section~\ref{SI-sec:additional-results} of the SI we present an additional analysis based on the dynamics of electrode charges, which approach smoothness for the system studied in Section~\ref{sec:diffusion}, consistent with the findings above.

These results thus suggest that there is a narrow window where one benefits from the electrostat over constant $E$.
This window is located at relatively small system sizes implying that using the stochastic description might be most beneficial in systems that are simulated, \textit{e.g.}, using more expensive potential energy surfaces (first-principles, or highly accurate machine-learning potentials). 
That being said, from a practical perspective, using the electrostat over constant-$E$ brings about only a marginal computational overhead, so using the complete description as a first choice is advisable, as there are no guarantees that the relevant system sizes are homogeneous across different systems and setups.
In particular, the system in question is a heavy and slow-moving, coarse-grained ionic liquid.
For other systems, for example those with explicit fast-vibrating bonds or with an explicit electronic structure, the stochastic noise of the electrostat may lead to a more pronounced difference between the ensembles at larger system sizes.

\section{\label{sec:Conclusion} Conclusion}

In this work, we derived the electrostat, a potential-control device that combines finite-field \ac{md} with
correct thermodynamical voltage fluctuations in capacitor-like systems.
The method we present represents a fully periodic, finite-field counterpart of the stochastic potentiostat and has been implemented in the \textsc{Electrode} package within the \textsc{Lammps} software.
From a broader perspective, this work demonstrates that thermodynamically robust voltage control can be achieved using a finite field in a periodic system, hence bridging finite-field and constant-voltage \ac{md}.

Our results lead to two central conclusions.
Firstly, building on the equivalence between the electrode-based (2D \ac{pbc}) and field-based (3D \ac{pbc}) ensembles, we find that the electrostat implementation is robust and exactly reproduces the expected physics.
Secondly, we found that for a broad range of systems, the methodologically simpler constant-$E$ approach yields various equilibrium properties indistinguishable from the physically rigorous electrostat.
An instructive parallel is that between isokinetic and canonical ensembles.
The reason why isokinetic \ac{md} is not routinely used to represent constant-temperature conditions is that equilibrium properties often deviate from canonical \ac{md} even for reasonably large systems.
However, unlike temperature control, voltage fluctuations affect the microscopic dynamics more indirectly through the electrode charge, which primarily acts as an electrostatic environment.

Our results put practical bounds on the importance of the electrostat in comparison to the constant-$E$ method and indicate that the stochastic fluctuations may become negligible for typical electrochemical systems with thousands of atoms.
This observation is consistent with results from classical \ac{md} models of similar size on the dielectric constant of water at interfaces with metals~\cite{stark_consistent_2026}.
However, for smaller systems the fluctuations in the potential can be crucial, as evidenced by our analysis of the size dependence of the charge fluctuations of the electrode for different system sizes. Simulations of proton-coupled electron transfer rates~\cite{sun_electrochemical_2026} in the literature seem to corroborate our assessment. 

A central computational advantage of the finite-field formulation of the electrostat is the fact that the charge accumulation at the electrode interfaces is imposed by the external field as opposed to the potentiostat approach in which charge states are imposed directly.
In this work, we have demonstrated the functionality on the simplest case of an empirical force field, and an extension of the electrostat to first-principles context is envisaged. This would include both  explicit electronic structure methods (namely, density-functional theory) and field-aware machine-learning force fields.

A first-principles electrostat implementation will require additional development, especially if it is to use the fully general \cref{eq:ou-process} rather than \cref{eq:ou-process-deissenbeck}.
The main problem to address is finding efficient ways of evaluating the total cell polarization $P$ and its time-derivative $j$ during propagation.
However, the benefit of this strategy is twofold: (1) it would allow us to overcome the limitations imposed on \cref{eq:ou-process-deissenbeck} and (2) it offers a description free of geometry-related quantities, such as $g$, $C_0$ or $j_\mathcal{E}$, as they all stem from the formal separation of the system into an electrode and an electrolyte part. This is particularly advantageous for a fully \textit{ab initio} setup, as separating the different parts of the system in this case becomes less well defined~\cite{mannino_first-principles_2026}.

Such developments will enable electrochemical simulations of charged interfaces not only with an unprecedented accuracy, but also allow us to extend to reactive systems and treat electrochemical phenomena under this unified framework.

\appendix

\section{Electrostatic equivalence between ensembles}
\label{app:theory-equivalence}

The relevant system geometries (shown in \cref{fig:cell-definition}) consist of an electrolyte placed between two planar conducting electrodes; in the following, the normal direction is chosen to be collinear with the $z$ Cartesian axis. 

We understand the electrodes to be ideal conductors and, thus, equipotentials~\cite{griffiths_introduction_1989}.
Such systems can be described in two distinct ways that were numerically shown to be equivalent~\cite{dufils_simulating_2019, dufils_computational_2021}. 
On the one hand, one can use 2D \acp{pbc} in $(x, y)$, keeping the outer electrode surfaces facing vacuum, and control the potential drop between the electrodes via the \ac{cpm} or \ac{ccm} methods (\cref{fig:cell-definition}A and B).
On the other hand, the system can be closed in 3D \acp{pbc} where the electrodes merge into one, leaving only two distinct electrolyte-electrode interfaces, and the electrochemical conditions are imposed by the action of finite fields (\cref{fig:cell-definition}C and D).
In previous works~\cite{dufils_simulating_2019}, the case \cref{fig:cell-definition}A was referred to as the electrolyte-centered system and the case~\cref{fig:cell-definition}C the conductor-centered system. 
Note that in the latter, centering of the unit cell is arbitrary due to the use of periodicity, as also demonstrated in \cref{fig:layer-charges}.

\subsection{Equivalence between \ac{cpm} and constant-$E$ simulations}

We set the stage by first discussing the case of 2D \ac{pbc}.
Here, imposing a voltage drop $\phi^*$ between the two electrodes is equivalent to fixing a Dirichlet boundary condition~\cite{griffiths_introduction_1989} 
\begin{equation}
\begin{split}
    &\phi(0) = \phi[(L - L')/2] = 0
    \\
    &\phi(L) = \phi[(L + L')/2] = \phi^*
\end{split}
\end{equation}
for the solution of Poisson equation
\begin{equation}
\label{eq:poisson}
    \bm{\nabla}^2\phi(\mathbf{r})
    =
    -\frac{1}{\varepsilon_0} \rho(\mathbf{r}),
\end{equation}
where $\phi(\mathbf{r})$ is the electrostatic potential, $\varepsilon_0$ is the vacuum permittivity, and $\rho(\mathbf{r})$ is the charge density within the cell.

On the electrostatically non-trivial region between the plates ($z \in [(L - L')/2, (L + L')/2]$), we invoke Lorentz averaging~\cite{griffiths_introduction_1989} and split the solution into two parts owing to the linearity of the Eq.~\ref{eq:poisson}
\begin{equation}
    \phi(\mathbf{r})
    =
    \phi_\mathrm{linear}(z) 
    + 
    \phi_\mathrm{grounded}(\mathbf{r}).
\end{equation}
The linear (macroscopic) term is defined as
\begin{equation}
    \phi_\mathrm{linear}(z)
    =
    \frac{\phi^*}{L'}z
    \label{eq:linear-potential}
\end{equation}
and therefore is responsible for the compliance of the solution with the Dirichlet boundary conditions.
Since the required boundary condition is fully accounted for, the ``grounded'' term $\phi_\mathrm{grounded}(\mathbf{r})$ is the solution of Eq.~\ref{eq:poisson} with the corresponding homogeneous boundary condition
\begin{equation}
\begin{split}
    &\phi_\mathrm{grounded}(0) = \phi_\mathrm{grounded}[(L - L')/2] = 0
    \\
    &\phi_\mathrm{grounded}(L) = \phi_\mathrm{grounded}[(L + L')/2] = 0
\end{split}
\end{equation}
and accounts for all the microscopic contributions arising from the spatial dependence of $\rho(\mathbf{r})$ under grounded conditions.
We choose zero as the value of the grounding potential without any loss of generality.

Making the $z$-direction periodic results in merging the electrodes into a single one.
First, in the absence of external fields, the electrostatic potential in the now fully periodic cell (which we will call $\phi_\mathrm{periodic}(\mathbf{r})$) must conform to the \ac{pbc} and fulfill
\begin{equation}
\begin{split}
    &\phi_\mathrm{periodic}(0) = \phi_\mathrm{periodic}[(L - L')/2] = 0
    \\
    &\phi_\mathrm{periodic}(L) = \phi_\mathrm{periodic}[(L + L')/2] = 0.
\end{split}
\end{equation}
Clearly, in the region between the plates and inside the electrodes, this potential corresponds exactly to $\phi_\mathrm{grounded}$ as the periodic images along $z$ are perfectly shielded from each other by the conducting electrodes and the charge distribution inside is otherwise identical.

Finally, once an external homogeneous field of magnitude equal to that in \cref{eq:e-star} along $z$ is imposed, this offsets the value of $\phi_\mathrm{periodic}$ at the \textit{cell} boundaries by $\phi^*$.
Each electrode adapts its interfacial charge distribution to this field to assumes a constant potential value throughout its whole volume~\cite{landau_electrodynamics_1984, griffiths_introduction_1989}.
The entire potential drop happens in the inner region between the plates, contributing a $-\phi^* / L'z$ term to the potential in that region.
This corresponds to $\phi_\mathrm{linear}(z)$ of \cref{eq:linear-potential} and the magnitude of the physical field established in the dielectric volume is correctly $E^*/g$, with $g$ given by~\cref{eq:geo-term}.
As solutions of the Poisson equations are unique~\cite{griffiths_introduction_1989} and the equation is identical for both above systems on $(0, L)$, we conclude that both systems are electrostatically equivalent inside a unit cell and a \ac{cpm} simulation can be mimicked by the appropriate constant $E$-field.

If electrode atoms were allowed to move, $g$ would cease to adopt a single well defined value.
However, this is not a problem for the constant-$E$ method, since \cref{eq:e-star} does not depend on $g$: the formal equivalence between \ac{cpm} and the constant-$E$ method holds even with mobile electrodes, because each electrode is an equipotential, irrespective of its immediate deformation.

\subsection{Equivalence between \ac{ccm} and constant-$D$ simulations}

A similar equivalence argument can be built for the open-circuit boundary condition when the system is disconnected from the electronic reservoir.
In the \ac{ccm} system, amount of total charge $Q^*$ deposited (with opposite signs) on the two electrode plates is imposed.
For the system geometries in question, $Q^*$ is linked to the external displacement field $D$ through Gauss's law 
\begin{equation}
    Q^* 
    =
    AD^*,
    \label{eq:gauss}
\end{equation}
as the $D$-field inside the cell is constant, and therefore, its flow through the cell boundary equals its flow through the physical electrode/electrolyte interface.
Hence, in the constant-$D$ ensemble, $Q^*$ deposited on the electrode interfaces can be controlled by imposing $D^*$.
From the definition of the electric displacement field (\cref{eq:def-D}), it becomes apparent that $E$ (and correspondingly $\phi$) can fluctuate when $D^*$ is fixed. 
Fixing $D^*$ is most easily achieved by imposing a homogeneous field $E = \varepsilon_0^{-1}(D^* - P)$ that varies at each step of the dynamics following the changes in $P$ due to the motion of the electrolyte and the interfaces.~\cite{stengel_electric_2009}

As discussed above in Section~\ref{sec:Theory}, the practical force-field implementation using $P_\mathcal{E}$ contains the effective $D$-field magnitude $gQ^*/A$.
It is thus apparent that knowing the value of $L'$ is required in this case, if one wants to relate $D^*$ to $Q^*$.
In practice, it can be obtained through $L' = \varepsilon_0 A / C_0$~\cite{zubeltzu_redefining_2025}.
The equation for $C_0$ depends on the method employed: within the \textsc{Electrode} package, it can be conveniently calculated from the electrode-electrode interaction matrix $\mathbf{A}$ (\cf{} \cref{eqn:Uele} and \cref{SI-eqn:C0} of the SI). 

Since the effective $D$-field depends on $L'$, letting the electrode atoms move breaks the formal equivalence between \ac{ccm} and the constant-$D$ method.
In principle, one could apply a constant $D^*$ in the same way as described above, but this would not correspond to the constant-charge ensemble, since the shape of the interface becomes geometry-dependent and the charge would fluctuate.
Employing a fluctuating value of $D$ to keep $Q^*$ constant is equally non-trivial.
A practical workaround would rely on effective constant parameters, for instance, an average $L'$ value.

Moreover, for \textit{ab initio} \ac{md}, the distinction between electrode and electrolyte charges is not straightforward and the definition of $L'$ becomes more intricate~\cite{mannino_first-principles_2026}.
Using the full $P$ instead of $P_\mathcal{E}$ would require additional methodology developments to overcome the inherent self-consistency issue of $P$ depending on an \textit{a priori} unknown $E$: this is similar in spirit to the electrostat case with $j$ rather than $j_\mathcal{E}$ discussed above in Section~\ref{sec:Conclusion}.

\section{Derivation of the electrostat equation of motion}
\label{app:theory-eom}

In the following, we derive an expression for
$\partial H(x; E^*)/\partial D$ from an \textit{ab initio}
perspective, where $H = \min_\psi \mathcal{H}$ and
$\mathcal{H}(x; \psi, E^*)$ is defined in \cref{eq:legendre}.
At fixed nuclear positions and momenta and fixed $D$, the minimizing electronic
state $\psi_D$ satisfies
\begin{equation}
    \frac{\delta \mathcal{H}}{\delta \psi}\bigg\vert_{\psi_D}
    =
    \frac{\delta H_0}{\delta \psi}
    - \Omega E \frac{\delta P}{\delta \psi}
    = 0,
    \label{eq:SSV-minim}
\end{equation}
where we have suppressed all arguments other than $\psi$ and introduced the
internal field
\begin{equation}
    E = \frac{D - P(\mathbf{r}; \psi)}{\varepsilon_0}.
    \label{eq:internal-field}
\end{equation}
Strictly, the minimization is constrained to a fixed number of electrons, so the right-hand side of \cref{eq:SSV-minim}
should contain the corresponding Lagrange-multiplier terms, \textit{e.g.}, the electronic chemical potential when $\psi$ is the electron density or a set of \ac{cpm} charges.
Since these constraints do not depend on $D$, they do not affect what follows.

Because $\mathcal{H}$ is stationary with respect to $\psi$ at $\psi_D$, the
Hellmann--Feynman theorem gives
\begin{equation}
\begin{split}
    \frac{\partial H}{\partial D}
    &=
    \frac{\partial \mathcal{H}}{\partial D}\bigg\vert_{\psi_D} \\
    &=
    \frac{\Omega}{\varepsilon_0}\left[D - P(\mathbf{r}; \psi_D)\right] - \Omega E^* \\
    &=
    \Omega \left(E - E^*\right),
    \label{eq:dH-dD}    
\end{split}
\end{equation}
with $E$ evaluated at $\psi_D$.

We note that, at fixed nuclear positions, \cref{eq:SSV-minim} is identical to
the electronic stationarity condition of $H_E$ (\cref{eq:ssv}) with $E^*$
replaced by the instantaneous internal field $E$, determined self-consistently
with $\psi_D$ through \cref{eq:internal-field}. The nuclear forces coincide as
well, since
\begin{equation}
    - \bm{\nabla}_\mathbf{r}H
    = 
    - \bm{\nabla}_\mathbf{r}H_0
    + 
    \Omega E \bm{\nabla}_\mathbf{r}P
 \end{equation}
which is the force derived from $H_E$ with $E^* \to E$.
This is what allows the electrostat to be realized by applying the constant-$E$
Hamiltonian with a field updated at every time step.

With these relations in hand, we start the derivation of the electrostat equation from the thermal phase-space probability distribution in \cref{eq:ensemble}
\begin{equation*}
    \rho(x, t)
    \propto
    e^{-\beta H(x; E^*)},
\end{equation*}
where $x=(\mathbf{r},\mathbf{p},D)$ with $x_D\equiv D$. 
Equivalent results can be obtained with the $E$-based formulation as demonstrated in Section~\ref{SI-sec:es-derivation} of the SI.
The time evolution of phase-space probability distributions is generally given by the Fokker--Planck equation~\cite{gardiner_hanbook_1985,kloeden_numerical_1999}
\begin{equation}
\begin{split}
    \frac{\partial \rho(x, t)}{\partial t}
    &=
    -\sum_{i} \frac{\partial}{\partial x_i} \left[ \mu_i(x, t)\rho(x, t) \right] \\
    &+ \sum_{i}\sum_{j} c_{ij} \frac{\partial^2 \rho(x, t)}{\partial x_i \partial x_j} \\
    &\equiv
    -\bm{\nabla} \cdot \bm{\mathcal{J}}(x, t)
\end{split}
\end{equation}
where the first term on the right represents its drift with drift vector $\bm{\mu}$ and the second term its diffusion (or spreading) with a diffusion matrix $\mathbf{c}$, where we have assumed constant $c_{ij}$.
Note that in the last line of the above equation, we have written the right-hand side as a divergence of a phase-space current $\bm{\mathcal{J}}$ in the form of a continuity equation.
The indices $i$ and $j$ are understood to run through all $6N + 1$ degrees of freedom, including the field stochastic variable.
In our case, the situation simplifies for two reasons.
First, we require our distribution to be stationary, so
\begin{equation}
    \frac{\partial \rho(x, t)}{\partial t}
    =
    0.
\end{equation}
Second, because we are describing conservative dynamics of the atomistic system (and even with coupling, \eg{}, a stochastic thermostat to the electrostat, the following argument remains the same), only the field degree of freedom has non-zero diffusion, \ie{}, $c_{ij} = 0$ for all elements but the single diagonal one, which we call $c_{D}$.
This gives us
\begin{equation}
\begin{split}
    &\sum_{i \neq D} \frac{\partial}{\partial x_i} \left[ \mu_i(x)\rho(x) \right] \\
    &+ \frac{\partial}{\partial D} \left[ \mu_D(x)\rho(x) \right] \\
    &- c_D \frac{\partial^2 \rho(x)}{\partial D^2} \\
    &= 0.
\end{split}
\end{equation}
The first sum corresponds to the terms of the divergence of $\bm{\mathcal{J}}$ due to the positions and momenta, which vanishes by virtue of Liouville's theorem~\cite{tuckerman_statistical_2010} and because $\rho$ is a function of $H$, which is conserved at constant $D$.

Hence, the Fokker--Planck equation reduces to only the $D$-derivative terms and the only way to ensure stationarity and normalizability of the probability in the remaining one-dimensional problem is letting the residual current vanish altogether by equating the drift and the diffusion terms
\begin{equation}
    \mu_D(x) \rho(x) 
    =
    c_D \frac{\partial \rho(x)}{\partial D}.
\end{equation}
Now, we evaluate the derivative term using \cref{eq:dH-dD}, which yields
\begin{equation}
    \mu_D
    =
    -c_D \beta \Omega (E - E^*).
\end{equation}
We identify
\begin{equation}
    \tau
    \equiv 
    \frac{1}{c_D \beta \Omega},
\end{equation}
where $\tau$ is a free parameter that governs the time scale of the electrostat (this is analagous to the parameter $\gamma$ in the derivation of the Langevin thermostat~\cite{bussi_accurate_2007}; however, note that $\tau$ does not yet have the dimension of time, as it relates $D$ and $E$ at this stage and thus contains an extra factor with the dimension of permittivity).

The Fokker--Planck dynamics of the probability distribution has a direct correspondence with an It\^{o} \acs{sde} for the dynamical trajectories of the fields themselves that sample the distribution~\cite{gardiner_hanbook_1985,kloeden_numerical_1999}.
For $D$, this reads
\begin{equation}
    \dd D
    =
    \mu_D \ \dd t + \sqrt{2c_D} \ \dd W_t
    =
    - \frac{1}{\tau} (E - E^*) \ \dd t 
    + 
    \sqrt{\frac{2}{\tau}\frac{k_\mathrm{B}T}{\Omega}} \ \dd W_t,
\end{equation}
where $W_t$ is a Wiener process.
At this point, we aim to eliminate $D$ in favor of $E$ by invoking its definition in \cref{eq:internal-field}, yielding
\begin{equation}
    \dd E
    =
    -\frac{1}{\tau \varepsilon_0} (E - E^*) \ \dd t 
    - \varepsilon_0^{-1} \dd P 
    + \sqrt{\frac{2}{\tau \varepsilon_0^2}\frac{k_\mathrm{B}T}{\Omega}} \ \dd W_t
\end{equation}
which transforms readily into
\begin{equation}
    \dd E
    =
    -\frac{1}{\tau_0} (E - E^*) \ \dd t 
    - \varepsilon_0^{-1} j \ \dd t
    + \sqrt{\frac{2}{\tau_0}\frac{k_\mathrm{B}T}{\varepsilon_0\Omega}} \ \dd W_t    
\end{equation}
where we relabeled $\tau \varepsilon_0 = \tau_0$ to arrive to the exact from of \cref{eq:ou-process}; $\tau_0$ has a dimension of time.

\section{Equivalence between the potentiostat and the electrostat}
\label{app:theory-es-vs-ps}

\subsection{Linearly polarizing ideal metal capacitor}

First, we derive the electronic susceptibility $\chi_0$ of an empty capacitor with an ideal metal electrode kept under $E^*$ and under the usual 3D \ac{pbc}.
As discussed in Appendix~\ref{app:theory-equivalence}, the physical field established after polarizing the electrode is zero inside the electrode ($E_\mathrm{metal}$) and a homogeneous $E^*/g$ in the vacuum ($E_\mathrm{vacuum}$).
The surface charge density $\sigma$ deposited on the electrode surface is given by the Gauss's law~\cite{griffiths_introduction_1989}
\begin{equation}
    \frac{\sigma}{\varepsilon_0}
    =
    E_\mathrm{vacuum} - E_\mathrm{metal} 
    =
    \frac{E^*}{g}.
\end{equation}
At the same time, the dipole moment induced by $E^*$ is
\begin{equation}
    \Delta \mu 
    =
    \sigma A(L - L'),
\end{equation}
where $L - L'$ is the thickness of the electrode slab.
Equating $\sigma$ in both expressions yields
\begin{equation}
    \Delta P
    =
    \frac{\Delta \mu}{\Omega}
    =
    \varepsilon_0 \left( \frac{1}{g} - 1 \right) E^*,
\end{equation}
in which we identify
\begin{equation}
    \chi_0
    \equiv
    \frac{1}{g} - 1.
\end{equation}
Note that the susceptibility becomes a purely geometric property under the ideal metal assumption.

At this point, we introduce the dielectric, which is assumed to be electronically non-polarizable but might carry a fixed, intrinsic polarization $P'_\mathcal{E} = \mu_\mathcal{E}/\Omega'$; note that we are using $\Omega' = AL'$ to normalize the polarization for now.
$P'_\mathcal{E}$ introduces an additional polarization of the metal under \acp{pbc}.
To quantify it, let us imagine for an instant the periodic array of polarized dielectric blocks with the metal slabs removed and under a macroscopic $E^*=0$ condition and ask what is the electric field $E_\mathrm{out}$ in the vacuum regions: this is precisely the field that the electrode has to screen.
\acp{pbc} require
\begin{equation}
    \Delta \phi
    =
    0
    =
    - \int_0^L \dd z \ E(z)
    =
    L' E_\mathrm{in}
    +
    (L - L') E_\mathrm{out}.
\end{equation}
Since there is no free charge inside this arrangement, $D(z)$ has no divergence and is thus constant over the whole unit cell and we have using \cref{eq:def-D}
\begin{equation}
    L' \frac{D - P'_\mathcal{E}}{\varepsilon_0}
    =
    (L' - L)\frac{D}{\varepsilon_0}
\end{equation}
which implies
\begin{equation}
    E_\mathrm{out}
    = 
    \frac{gP'_\mathcal{E}}{\varepsilon_0}
    =
    \frac{P_\mathcal{E}}{\varepsilon_0}.
\end{equation}

Putting it all together, the total polarization of the ideal metal capacitor filled with dielectric carrying an intrinsic polarization $P_\mathcal{E}$ but electronically non-polarizable is
\begin{equation}
    P
    =
    P_\mathcal{E}
    +
    \varepsilon_0\chi_0E_\mathrm{out} 
    +
    \varepsilon_0\chi_0E^*
    =
    \frac{P_\mathcal{E}}{g} 
    +
    \varepsilon_0 \left( \frac{1}{g} - 1 \right)E^*,
\end{equation}
as stated in \cref{eq:linear-polarization}, which combined with \cref{eq:ou-process} gives \cref{eq:ou-process-deissenbeck}.

\subsection{Relation to the original thermopotentiostat equation}
The authors of the original stochastic (thermo)potentiostat implementation~\cite{deisenbeck_dielectric_2021} state the original  equation of motion in terms of the total electrode charge $Q$ which reads
\begin{equation}
\begin{split}
    Q(t + \Delta t)
    &=
    Q(t) - C_0[\phi(t) - \phi^*] \left(1 - \alpha\right) \\
    &+
    \mathcal{N}(t) \sqrt{k_\mathrm{B}TC_0\left(1 - \alpha^2 \right)}.    
\end{split}
\label{eq:eom-q}
\end{equation}
Eq.~\ref{eq:eom-q} does not contain the polarization explicitly.
We now show that it enters implicitly through $\phi$, as in \cref{eq:es-eom}, demonstrating the equivalence between the two approaches, within limits discussed below.

In Eq.~4 of  Ref.~\citenum{deisenbeck_dielectric_2021}, the authors state
\begin{equation}
    \dd\phi
    =
    \frac{1}{C_0} (\dd Q + \dd Q_\mathrm{p}),
    \label{eq:dphi}
\end{equation}
where $Q_\mathrm{p}$ is the ``polarization'' charge ($Q_\mathrm{p}=-P'_\mathcal{E}A$).
As a next step, $C_0^{-1} \dd Q = f \dd t$ is introduced and expressed through the Ornstein--Uhlenbeck process to obtain  \cref{eq:eom-q}.

We first divide the whole equation by $C_0$:
\begin{equation}
\begin{split}
    \frac{1}{C_0}Q(t + \Delta t)
    &=
    \frac{1}{C_0}Q(t) - [\phi(t) - \phi^*] \left(1 - \alpha \right) \\
    &+
    \mathcal{N}(t) \sqrt{\frac{k_\mathrm{B}T}{C_0}\left(1 - \alpha^2 \right)}
\end{split}
\end{equation}
and fill in for $Q$ from \cref{eq:dphi} to get
\begin{equation}
\begin{split}
    \phi(t + \Delta t) - \frac{1}{C_0}Q_\mathrm{p}(t + \Delta t)
    &=
    \phi(t) - \frac{1}{C_0}Q_\mathrm{p}(t) \\
    &- [\phi(t) - \phi^*] \left(1 - \alpha \right) \\
    &+
    \mathcal{N}(t) \sqrt{\frac{k_\mathrm{B}T}{C_0}\left(1 - \alpha^2 \right)}.    
\end{split}
\end{equation}
This can be simplified to
\begin{equation}
\begin{split}
    \phi(t + \Delta t) - \phi^*
    &=
    \frac{1}{C_0} [Q_\mathrm{p}(t + \Delta t) - Q_\mathrm{p}(t)] \\
    &+ \alpha[\phi(t) - \phi^*] \\
    &+ \mathcal{N}(t) \sqrt{\frac{k_\mathrm{B}T}{C_0}\left(1 - \alpha^2\right)}.    
\end{split}
\end{equation}
The first term on the right is the change in potential due to the change in polarization stemming from the motion of the electrolyte.
This is easily shown using
\begin{equation}
\begin{split}
    \frac{1}{C_0} [Q_\mathrm{p}(t + \Delta t) - Q_\mathrm{p}(t)]
    &=
    -\frac{L'}{\varepsilon_0} [P'_\mathcal{E}(t + \Delta t) - P'_\mathcal{E}(t)] \\
    &=
    -\varepsilon_0^{-1} L \frac{\dd P_\mathcal{E}(t)}{\dd t} \Delta t \\
    &=
    -\varepsilon_0^{-1} L j_\mathcal{E}(t) \Delta t 
    \label{eqn:j-from-Qp}
\end{split}
\end{equation}
Apart from the additional factor $L$ that comes from the fact that the potentiostat deals with the potential rather than $E$, \cref{eq:es-eom} and \cref{eq:eom-q} describe identical physics in their respective \acp{pbc}.

To complete the analogy, note that the potentiostat equation contains $\Delta t$ rather than the exact prefactor $\tau_0'( 1 - \alpha)$ that we have derived for the electrostat in the current term.
This is a viable approximation in the limit of small time steps $\Delta t \ll \tau_0'$, since expanding the exponential into a power series and truncating after the first order
\begin{equation}
\begin{split}
    \tau_0' \left( 1 - e^{-\frac{\Delta t}{\tau_0'}} \right)
    &=
    \tau_0' \left[ 1 - \left( 1 - \frac{\Delta t}{\tau_0'} \right) \right] + \mathcal{O}\left[\left(\frac{\Delta t}{\tau_0'}\right)^2\right] \\
    &\approx
    \Delta t.
    \label{eqn:exp-truncation}
\end{split}
\end{equation}

\section*{Supplementary material}

See the supporting information for more information on details of the employed computational workflows, details of the mathematical derivations, derivation of supporting identities, details of the \textsc{Lammps/Electrode} implementation of the finite-field methods and additional benchmarks and results supporting the claims of the main text.

\begin{acknowledgments}
The authors acknowledge insightful discussions with Gabor Cs\'anyi and William Baldwin. M.R. acknowledges funding by the European Research Council (ERC) Consolidator Grant (CoG) under the European Union’s Horizon Europe research and innovation programme (Grant agreement No. 101169761, Project Acronym QUADYMM).
We also thank Robert Mei{\ss}ner for inspiring us to look at dynamical properties under different conditions and for asking the right questions for us to clarify. 
This work was partially funded by the Deutsche Forschungsgemeinschaft (DFG, German
Research Foundation) under Germany's Excellence Strategy – EXC 3120/1 – 533771286.
\end{acknowledgments}

\section*{Author declarations}

\subsection*{Conflict of interest}

The authors have no conflict of interest to declare.

\subsection*{Author contributions}

Krystof Brezina: Conceptualization (supporting); Data curation (equal); Formal analysis (equal); Methodology (lead); Validation (equal); Visualization (equal); Writing -- original draft (equal)

Ludwig Ahrens-Iwers: Data curation (equal); Formal analysis (equal); Methodology (supporting); Software (lead); Validation (equal); Visualization (equal); Writing -- original draft (equal)

Mariana Rossi: Conceptualization (lead); Funding aquisition (lead); Methodology (supporting); Project administration (lead); Resources (lead); Supervision (lead); Writing -- original draft (equal)

\section*{Data availability}

The data supporting the findings of this study are available in Zenodo at \url{doi.org/10.5281/zenodo.23401158}.

\bibliography{main,more}

\end{document}



\title{Supplementary information for: Electrochemical Molecular Dynamics within the Finite-Field Framework}

\author{Krystof Brezina}
\affiliation{Max Planck Institute for the Structure and Dynamics of Matter, Luruper Chaussee 149, 22761 Hamburg, Germany}
\author{Ludwig Ahrens-Iwers}
\affiliation{Max Planck Institute for the Structure and Dynamics of Matter, Luruper Chaussee 149, 22761 Hamburg, Germany}
\author{Mariana Rossi}
\affiliation{Max Planck Institute for the Structure and Dynamics of Matter, Luruper Chaussee 149, 22761 Hamburg, Germany}
\affiliation{Yusuf Hamied Department of Chemistry, University of Cambridge, Lensfield Road, Cambridge CB2 1EW, U.K.}

\date{\today}
\maketitle

\input{acronyms.tex}

\section{Derivation of the equation of motion of the electrostat through $E$}
\label{sec:es-derivation}

As discussed in the main text, it is actually not the $E$-field but rather the $D$-field that is the natural independent variable to describe the fluctuation of free charge at the cell boundary due to thermal noise in the circuit.
However, the governing physics is clearly agnostic to which one of the field variables is used to derive the equation of motion and an identical one may be obtained directly using $E$ as the driving variable after a coordinate transformation. 
We show that the algebra becomes less straightforward since $E$ is not independent of the particle configuration, but it rather mixes the system and external degrees of freedom $E = E(D, \mathbf{r})$.
Note that both derivations address the closed-circuit ensemble.

We start by transforming the electrostat Hamiltonian $H$ (defined in \cref{main-eq:legendre} of the main text) as a function of the internal field $E$ rather than $D$ using \cref{main-eq:internal-field} of the main text as
\begin{equation}
    \widetilde{H}(y; E^*)
    =
    H_0(\mathbf{r}, \mathbf{p}) + \frac{\Omega \varepsilon_0}{2} E^2 - \Omega E^* [\varepsilon_0E + P(\mathbf{r})],
    \label{eq:HE}
\end{equation}
with $y \equiv (\mathbf{r}, \mathbf{p}, E)$. 
We understand again that it has been minimized with respect to the electronic structure $\psi$.
As in the main text, we define the ensemble distribution
\begin{equation}
    \widetilde{\rho}(y) \propto e^{-\beta \widetilde{H}(y; E^*)}.
\end{equation}

The physics does not change under the transformation $H \rightarrow \widetilde{H}$, and the canonical positions and momenta must remain the same as those obtained in the main text.
This is satisfied for the velocities in the $E$-based framework 
\begin{equation}
    \dot{r}_i
    =
    \left( \frac{\partial H}{\partial p_i}\right)_{D}
    =
    \left( \frac{\partial \widetilde{H}}{\partial p_i} \right)_{E}
    =
    \frac{\partial H_0}{\partial p_i}.
\end{equation}
Note that for clarity, we highlight which of the relevant variables are fixed in each term.
For the forces, the partial derivative $(\partial \widetilde{H}/\partial r_i)_E$ is not the physical force $(\partial H/\partial r_i)_D$, because fixing $E$ forces $D$ to change along with $r_i$ and therefore the direction of the derivative at $\mathrm{d}E = 0$ does not correspond to a purely atomic displacement.
Rather, we must choose the differential of $E$ 
\begin{equation}
    \mathrm{d}E
    =
    \left( \frac{\partial E}{\partial D} \right)_{r_i} \mathrm{d}D
    +
    \left( \frac{\partial E}{\partial r_i} \right)_{D} \mathrm{d}r_i
\end{equation}
that satisfies $\mathrm{d}D = 0$ to evaluate the forces, hence
\begin{equation}
    \mathrm{d}E 
    =
    \left( \frac{\partial E}{\partial r_i} \right)_{D} \mathrm{d}r_i.
\end{equation}
The change in $\widetilde{H}$ due to displacing $r_i$ and $E$ in this way then is
\begin{equation}
    \mathrm{d}\widetilde{H}
    = 
    \left( \frac{\partial \widetilde{H}}{\partial r_i} \right)_{E} \mathrm{d}r_i
    +
    \left( \frac{\partial \widetilde{H}}{\partial E} \right)_{r_i} \mathrm{d}E
    =
    \left( \frac{\partial \widetilde{H}}{\partial r_i} \right)_{E} \mathrm{d}r_i
    +
    \left( \frac{\partial \widetilde{H}}{\partial E} \right)_{r_i}
    \left( \frac{\partial E}{\partial r_i} \right)_{D} \mathrm{d}r_i,
\end{equation}
which rearranges into
\begin{equation}
    \left( \frac{\partial \widetilde{H}}{\partial r_i} \right)_{D}
    =
    \left( \frac{\partial H}{\partial r_i} \right)_{D}
    =
    \left( \frac{\partial \widetilde{H}}{\partial r_i} \right)_{E}
    +
    \left( \frac{\partial \widetilde{H}}{\partial E} \right)_{r_i}
    \left( \frac{\partial E}{\partial r_i} \right)_{D}
    =
    -\dot{p}_i.
\end{equation}
Note that in the left-hand side, the subscript $D$ denotes taking the derivative along the direction of constant $D$.

Evaluating the derivatives using \cref{eq:HE}, we obtain the explicit correction to the force
\begin{equation}
    \dot{p}_i
    =
    -\left(\frac{\partial \widetilde{H}}{\partial r_i}\right)_{E} + \Omega \left( \frac{\partial P}{\partial r_i} \right)_D (E - E^*).
\end{equation}
The presence of this correction term leads to the fact that the divergence of the position and momentum currents in the Fokker--Planck equation does not vanish fully.
 Performing the same manipulations as in the main text yields the following residue
\begin{equation}
    \frac{\partial}{\partial r_i} (\dot{r}_i \widetilde{\rho}) + \frac{\partial}{\partial p_i} (\dot{p}_i \widetilde{\rho})
    =
    - \beta \widetilde{\rho} \Omega (E - E^*) \frac{\partial P}{\partial r_i} \dot{r}_i,
\end{equation}
which sums over all $i$ to
\begin{equation}
    -\beta \widetilde{\rho} \Omega (E - E^*) \frac{\mathrm{d} P}{\mathrm{d} t}.
\end{equation}

The physical phase-space distribution remains stationary, so the overall probability current must vanish as it did before.
Hence, the residual term obtained above enters the detailed balance condition, which now reads
\begin{equation}
    \beta \widetilde{\rho} \Omega (E - E^*) \frac{\mathrm{d} P}{\mathrm{d} t} 
    - \frac{\partial}{\partial E}(\mu_E \widetilde{\rho})
    + c_E \frac{\partial^2 \widetilde{\rho}}{\partial E^2}
    = 
    0.
\end{equation}
We use
\begin{equation}
    \frac{\partial \widetilde{\rho}}{\partial E}
    =
    -\beta \widetilde{\rho} \varepsilon_0 \Omega (E - E^*) 
\end{equation}
to simplify the equation above to
\begin{equation}
    \frac{\partial}{\partial E}
    \left[
        - \varepsilon_0^{-1} \frac{\mathrm{d} P}{\mathrm{d} t} \widetilde{\rho}
        - \mu_E \widetilde{\rho}
        - c_E \beta \widetilde{\rho} \varepsilon_0 \Omega (E - E^*) 
    \right]
    = 
    0    
\end{equation}
and, using the same argument as in Appendix~\ref{main-app:theory-eom} of the main text, we obtain the drift of $E$ as
\begin{equation}
    \mu_E
    =
    - c_E \beta \varepsilon_0 \Omega (E - E^*) - \varepsilon_0^{-1} \frac{\mathrm{d} P}{\mathrm{d} t},
\end{equation}
where we further associate $\tau_0^{-1} = c_E \beta \Omega \varepsilon_0$.
Note that the drift already naturally contains the polarization term.
It allows us to build the SDE using the Fokker--Planck/It\^{o} correspondence theorem directly for $E$, which has the same form as the one derived in the main text
\begin{equation}
    \dd E
    =
    -\frac{1}{\tau_0}(E - E^*) \ \dd t - \varepsilon_0^{-1} j \ \dd t + \sqrt{\frac{2}{\tau_0} \frac{k_\mathrm{B}T}{\varepsilon_0\Omega}} \dd W_t
\end{equation}
as required.

\section{Solution of the driven It\^{o} SDE}
\label{sec:ito-solution}

The final step between \cref{main-eq:ou-process-deissenbeck} and \cref{main-eq:es-eom} in the main text is the exact integration of the former to a finite time step.
This \ac{sde} represents a driven Ornstein--Uhlenbeck process~\cite{gardiner_hanbook_1985} with an extra deterministic term
\begin{equation}
    \dd x(t)
    =
    -
    \frac{1}{\tau'_0}x(t) \ \dd t 
    - 
    j(t) \ \dd t 
    + 
    \sigma \ \dd W_t
\end{equation}
and may be solved using the method of variation of constants and applying identities of stochastic calculus~\cite{kloeden_numerical_1999}.
We start by solving the homogeneous deterministic differential equation
\begin{equation}
    \dd x(t)
    + 
    \frac{1}{\tau'_0}x(t)\  \dd t 
    = 
    0.
\end{equation}
This is solved readily by
\begin{equation}
    x(t) 
    = 
    Ce^{-\frac{t}{\tau'_0}},
\end{equation}
where $C$ is a real constant.
Now we vary the constant $C \rightarrow f(x, t) \implies f(x, t) = x(t) e^{\frac{t}{\tau'_0}}$ to extend to the full solution and apply It\^{o}'s lemma~\cite{kloeden_numerical_1999} (the stochastic generalization of the chain rule)
\begin{equation}
    \dd f(x, t)
    =
    \frac{\partial f(x, t)}{\partial t} \dd t 
    + 
    \frac{\partial f(x, t)}{\partial x} \dd x
    +
    \frac{1}{2} \frac{\partial^2 f(x, t)}{\partial x^2} (\dd x)^2.
\end{equation}
With that, we have
\begin{equation}
    \dd f(x, t)
    =
    \frac{1}{\tau'_0} e^{\frac{t}{\tau'_0}} x(t) \ \dd t
    +
    e^{\frac{t}{\tau'_0}} \left[
    -
    \frac{1}{\tau'_0}x(t) \ \dd t 
    - 
    j(t) \ \dd t
    +
    \sigma \ \dd W_t
    \right]
    =
    -e^{\frac{t}{\tau'_0}} j(t) \ \dd t
    +
    e^{\frac{t}{\tau'_0}}\sigma \ \dd W_t.
\end{equation}
Now, we can integrate from $t=0$ to $t=\Delta t$.
For the left-hand side, we have
\begin{equation}
    \int_{0}^{\Delta t} \dd f
    =
    f(\Delta t) - f(0)
    =
    x(\Delta t) e^{\frac{\Delta t}{\tau'_0}} - x(0).
\end{equation}
For the first term on the right-hand side, we must remember that within a simulation time step $\Delta t$, $j(t)$ remains constant and can write
\begin{equation}
    \int_{0}^{\Delta t} e^{\frac{t}{\tau'_0}} j(t) \ \dd t
    =
    j(0) \tau'_0 \left( e^{\frac{\Delta t}{\tau'_0}} - 1 \right).
\end{equation}
For the stochastic integral, we exploit the fact that the variance of a Wiener process is deterministic and use the corresponding tools of stochastic calculus, specifically an integration rule sometimes referred to as the It\^{o} isometry~\cite{kloeden_numerical_1999, oksendal_stochastic_1998}.
With it, we can evaluate the mean of the square of the integral in question
\begin{equation}
    \mu \left[\left( \int_{0}^{\Delta t} e^{\frac{t}{\tau'_0}}\sigma \ \dd W_t \right)^2\right]
    =
    \sigma^2 \int_{0}^{\Delta t} e^{\frac{2t}{\tau'_0}} \ \dd t
    =
    \sigma^2 \frac{\tau'_0}{2} \left( e^{\frac{2\Delta t}{\tau'_0}} - 1 \right),
\end{equation}
which corresponds to the variance of the sought stochastic integral. 
The integral itself can be realized through
\begin{equation}
    \int_{0}^{\Delta t} e^{\frac{t}{\tau'_0}}\sigma \ \dd W_t
    =
    \mathcal{N}\sqrt{\sigma^2 \frac{\tau'_0}{2}\left( e^{\frac{2\Delta t}{\tau'_0}} - 1 \right)},
\end{equation}
where $\mathcal{N}$ is a Gaussian random number sampled from a normal distribution with a zero mean and unit variance.
Putting everything together and multiplying the whole expression by $e^{-\frac{\Delta t}{\tau'_0}}$, we recover the above given update formula
\begin{equation}
    x(\Delta t)
    =
    x(0) e^{-\frac{\Delta t}{\tau'_0}}
    -
    j(0) \tau'_0 \left( 1 - e^{-\frac{\Delta t}{\tau'_0}} \right)
    +
    \mathcal{N}\sqrt{\sigma^2 \frac{\tau'_0}{2} \left( 1 - e^{-\frac{2\Delta t}{\tau'_0}} \right)}.
\end{equation}

\section{Difference between the electrostat and potentiostat}
\label{sec:eq-es-tps}

The electrostat equation (\cref{main-eq:es-eom}) contains an electrolyte current term
\begin{equation}
    E_j(t) = - \tau'_0\left( 1 - \alpha \right) \varepsilon_0^{-1} j_{\mathcal{E}}(t).
\end{equation}
In order to relate the electrostat to the original thermopotentiostat algorithm~\cite{deisenbeck_dielectric_2021}, one can modify the equation of motion in the following way: $E_j(t)$ is omitted and the field $E(t)$ in the equation of motion is substituted by $E'(t)$, evaluated with retained \acs{cpm} charges but after the update of electrolyte positions.
Since the electrode charges, and hence the immediate $D$-field, remain unchanged, one can use \cref{main-eqn:eff-D} to show that the reevaluation introduces a term that approximates $E_j(t)$:
\begin{equation}
    E'(t) - E(t) = - \varepsilon_0^{-1} \Delta P_\mathcal{E}(t) = - \Delta t \varepsilon_0^{-1} j_\mathcal{E}(t),
\end{equation}
specifically, by replacing $\tau'_0\left( 1 - \alpha \right)$ with $\Delta t$.
As we have shown in \cref{main-eqn:exp-truncation}, this is a valid first order approximation but differs for higher orders.
This is not changed by the multiplication with $\alpha = \mathrm e ^{-\Delta t/\tau'_0}$ in the equation of motion.
As we have discussed in Appendix~\ref{main-app:theory-es-vs-ps}, the potentiostat uses exactly this approximation, which makes it fully equivalent to the modified electrostat based on $E'(t)$ in this regard.

\begin{figure}[tb!]
    \includegraphics{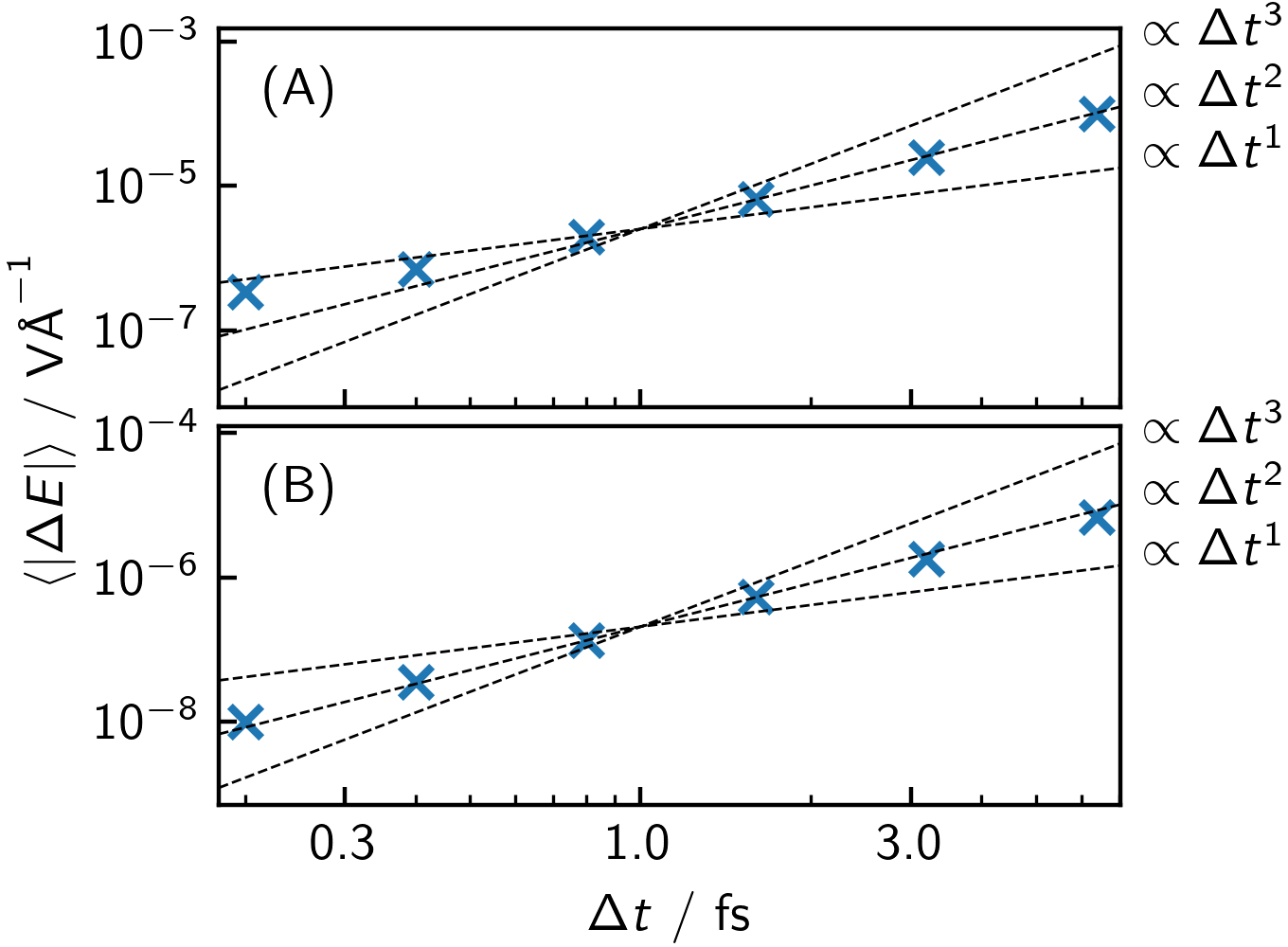}
    \caption{Average difference between the $E_j(t)$ and the equivalent term $\alpha \left[ E'(t) - E(t) \right]$ for a series of time steps.
    Panel (A) is for the capacitor with aqueous electrolyte and gold electrodes (\cf{} \cref{fig:commutation}), panel (B) for the ionic liquid system.
    Both systems show a trend of $\sim\Delta t^2$.}
    \label{fig:diff_j-term}
\end{figure}

We show the discrepancy between $E_j(t)$ and $\alpha \left[E'(t) - E(t)\right]$ explicitly by comparing the two terms for a range of time steps and two model systems in \cref{fig:diff_j-term}.
The difference shows the expected quadratic behavior with respect to the integration time step, suggesting larger deviations at time steps of several \unit{\femto\second}.

\section{Implementation details}
\label{sec:electrode}

\subsection{Constraints and vacuum capacitance}
\label{sec:implementation-efield}
In the constant-$E$, the task is to solve
\begin{equation}
    \mathbf A \mathbf q = \mathbf v + \mathbf{z}E^* - \mathbf b = \mathbf v' - \mathbf b
    \label{eqn:cpm}
\end{equation}
for $\mathbf q$, where we define an effective potential $\mathbf v' \equiv \mathbf v + \mathbf{z}E^* $ to simplify the implementation.
Using this definition, constant-$E$ is similar to the \ac{cpm}.
However, it differs in the use of $\mathbf v'$ instead of $\mathbf v$, as well as in the \acp{pbc} used for $\mathbf A$ and $\mathbf b$.
Further, there is only one electrode, and therefore the values of the potential $v_i = v_0$ all take the same value.
The potential needs to be continuous throughout the electrode, \ie{}, if electrode atoms cross a periodic boundary, an offset $\pm L$ is added to $z_i$.

For constant-$E$, the potential is written as $v'_i = E^* z_i + v_0$.
A new matrix
\begin{equation}
    \mathbf H \in \mathbb{R}^{N\times 2}\;, \quad H_{iJ} = \left(-1\right)^J \frac{z_i}{L} + \frac{1}{2}
    \label{eqn:H}
\end{equation}
is defined, allowing us to write the potential as
\begin{equation}
    \mathbf v' = \mathbf H \gw v \quad \mathrm{with} \quad \gw v = 
    \begin{pmatrix}
        -E^* L/2 + v_0 \\
        E^* L/2 + v_0 
    \end{pmatrix}.
    \label{eqn:v-tilde}
\end{equation}
Other choices of $\mathbf H$ and $\gw{v}$ are possible, \eg{} $\gw{v}^\mathrm T = (E^* L, v_0)$.
The presented form has the property, that an offset $v_0$ to $\gw{v}$ leads to an equal offset in $\mathbf v'$ without affecting the applied field.
Further, one can define an indicator matrix \cite{ahrens-iwers_electrode_2022}
\begin{equation}
    \mathbf G \in \mathbb{R}^{N\times n} \;, \quad G_{iJ} = 
    \begin{cases}
        1\; \text{if atom $i$ in electrode $J$}\\
        0\; \text{else}
    \end{cases}
\end{equation}
which assigns each of the $N$ electrode atoms to one of the $n$ electrodes.
By applying $\mathbf G^\mathrm T$ to \cref{eqn:cpm}, we obtain an expression for the electrode charges
\begin{equation}
    \gw q =  \mathbf G^\mathrm T \mathbf q = \gw C  \gw v - \mathbf G^\mathrm T \mathbf A^{-1} \mathbf b,
    \label{eqn:q-tilde}
\end{equation}
with the capacitance matrix $\gw{C} = \mathbf G^\mathrm T \mathbf A^{-1} \mathbf H \in \mathbb R^{n\times 2}$.
Employing this equation, one can compute the required offset potential
\begin{equation}
    v_0 = \frac{Q_0 - \sum_I \left(\gw{C} 
    \begin{pmatrix}
        -E^* L/2\\
        E^* L/2
    \end{pmatrix}
    - \mathbf G^\mathrm T \mathbf A^{-1} \mathbf b
\right)_I }{\sum_{IJ} \tilde{C}_{IJ}},
    \label{eqn:v0}
\end{equation}
for a given total charge $Q_0$.
For the constant-$E$ method, one can apply a field without separating the electrode atoms into different groups for each electrode, \ie{}, the equations above can be simplified by inserting $\mathbf G = (1,\ldots, 1)^\mathrm T$.

If one wishes to calculate the vacuum capacitance of a model system with two electrodes, one can set $n=2$ and solve
\begin{equation}
    \begin{pmatrix}
        -Q \\
        Q
    \end{pmatrix}
    = \gw{C}
    \begin{pmatrix}
        -\Delta v/2 + v_0 \\
        \Delta v/2 + v_0
    \end{pmatrix}
    \label{eqn:Q}
\end{equation}
for $Q/\Delta v$, yielding
\begin{equation}
    C_0 = \frac{\mathrm{det}\;\gw{C}}
        {\sum_{IJ} \tilde{C}_{IJ}}.
    \label{eqn:C0}
\end{equation}
The equations for the offset potential and the vacuum capacitance in \cref{eqn:v0,eqn:C0} are the same as for the \ac{cpm}~\cite{scalfi_charge_2020, ahrens-iwers_electrode_2022}.
The crucial difference lies in the expression for the potential in \cref{eqn:v-tilde} and the definition of the capacitance matrix $\gw C$ using $\gw H$.
This matrix is not symmetric, in contrast to the capacitance matrix in the \ac{cpm}.
\subsection{Constant $D$-field}
\label{sec:dfield}

For the constant $D$-field method, the applied electric field
\begin{equation}
    E(t)
    = 
        \varepsilon_0^{-1} \left[gD^* - P_\mathcal{E}(t)\right]
\end{equation}
is calculated from the $D$-field and the electrolyte polarization $P_\mathcal{E}(t)$.
There are two approaches to calculate $P_\mathcal{E}(t)$.
The first is to directly calculate the electrolyte dipole, \ie{}, $P_\mathcal{E}(t) = \Omega^{-1} \sum_{i \in \mathcal{E}} q_i z_i(t)$.
In the second approach, one exploits the fact that the electrode-electrolyte interaction $-\mathbf b$ can be interpreted as the potential due to the electrolyte.
Thus, one can define a group wise electrolyte potential $\gw v_b = -\gw C^{-1} \mathbf G^\mathrm T \mathbf A^{-1} \mathbf b$ and calculate the polarization $P_\mathcal{E} = \varepsilon_0\Delta \tilde v_b / L$ from the difference between its two values.
Alternatively, on can fit a linear function $-b_i = \varepsilon_0^{-1} P_\mathcal{E} z_i - b_0$ and obtain the polarization as the slope of this fit.
All presented options yield virtually the same results and we have decided for the first approach in our implementation for simplicity.

\subsection{Symmetric splitting of integrator}
\label{sec:symmetric-propagator}
The update to the $E$-field
\begin{equation}
\begin{split}
    \Delta E(t + \Delta t)
    =
    &\Delta E(t) e^{-\frac{\Delta t}{\tau'_0}} \\
    &-
    \tau'_0 \left( 1 - e^{-\frac{\Delta t}{\tau'_0}}\right) \varepsilon_0^{-1} j_\mathcal{E}(t) \\
    &+
    \mathcal{N}(t) \frac1L \sqrt{\frac{k_\mathrm{B}T}{C_0} \left( 1 - e^{-\frac{2\Delta t}{\tau'_0}}\right)}  
    \label{eq:full-eom}
\end{split}
\end{equation}
with $\Delta E(t) = E(t) - E^*$,
comprises a dissipation term, an electrolyte current term and a fluctuation term.
In the asymmetric B-ES-A-B splitting, the update is performed after the half step update of velocities (B) and full step update of positions (A).

In the symmetric B-ES-A-ES-B splitting, the $E$-field is propagated by $\Delta t/2$ before and after positions are updated.
Since the update of positions does not directly depend on the $E$-field, both steps can be performed after the update to positions for convenience.
The update then becomes
\begin{equation}
\begin{split}
    \Delta {E}(t + \Delta t)
    &=
     \left[  \Delta E(t) e^{-\frac{\Delta t}{2\tau'_0}} 
    - \tau'_0 \left( 1 - e^{-\frac{\Delta t}{2\tau'_0}}\right) \varepsilon_0^{-1} j_\mathcal{E}(t) 
    + \mathcal{N}(t) \frac1L \sqrt{\frac{k_\mathrm{B}T}{C_0} \left( 1 - e^{-\frac{\Delta t}{\tau'_0}}\right)}  \right] e^{-\frac{\Delta t}{2\tau'_0}} \\
    &\quad-
    \tau'_0 \left( 1 - e^{-\frac{\Delta t}{2\tau'_0}}\right) \varepsilon_0^{-1} j_\mathcal{E}(t) 
    + \mathcal{N}(t+\Delta t/2) \frac1L \sqrt{\frac{k_\mathrm{B}T}{C_0} \left( 1 - e^{-\frac{\Delta t}{\tau'_0}}\right)}  \\
    &=
    \Delta E(t) e^{-\frac{\Delta t}{\tau'_0}} 
    -  \tau'_0 \left( 1 - e^{-\frac{\Delta t}{\tau'_0}}\right) \varepsilon_0^{-1} j_\mathcal{E}(t) \\
    &\quad + \left[e^{-\frac{\Delta t}{2\tau'_0}} \mathcal{N}(t) + \mathcal{N}(t+\Delta t/2) \right] \frac1L \sqrt{\frac{k_\mathrm{B}T}{C_0} \left( 1 - e^{-\frac{\Delta t}{\tau'_0}}\right)}.
    \label{eq:half-step-eom}
\end{split}
\end{equation}
This equation yields the same expression as in \cref{eq:full-eom} for the first two terms, whereas the noise term uses two random numbers and is therefore different.

\begin{figure}
    \begin{minipage}{0.6\textwidth}
        \includegraphics{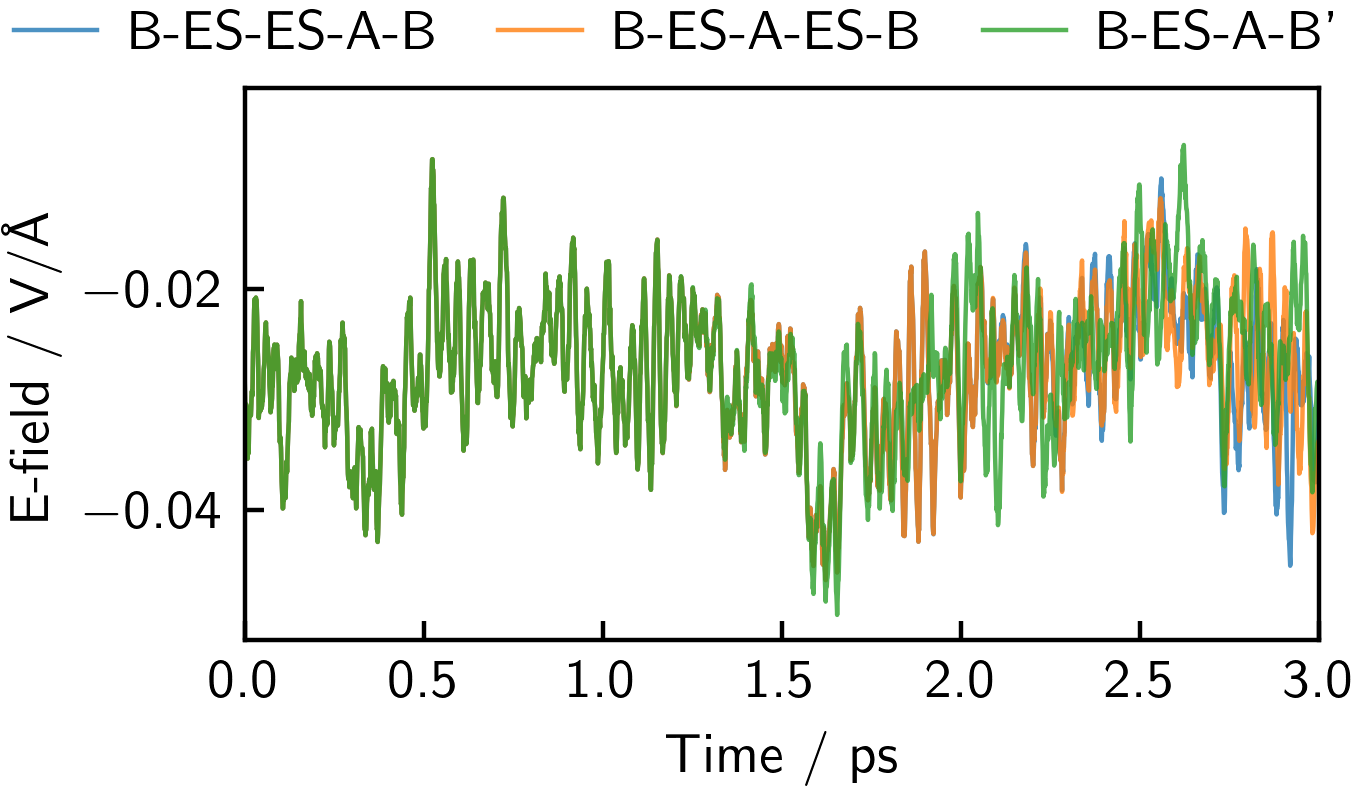}
    \end{minipage}
    \begin{minipage}{0.35\textwidth}
        \includegraphics[width=\linewidth]{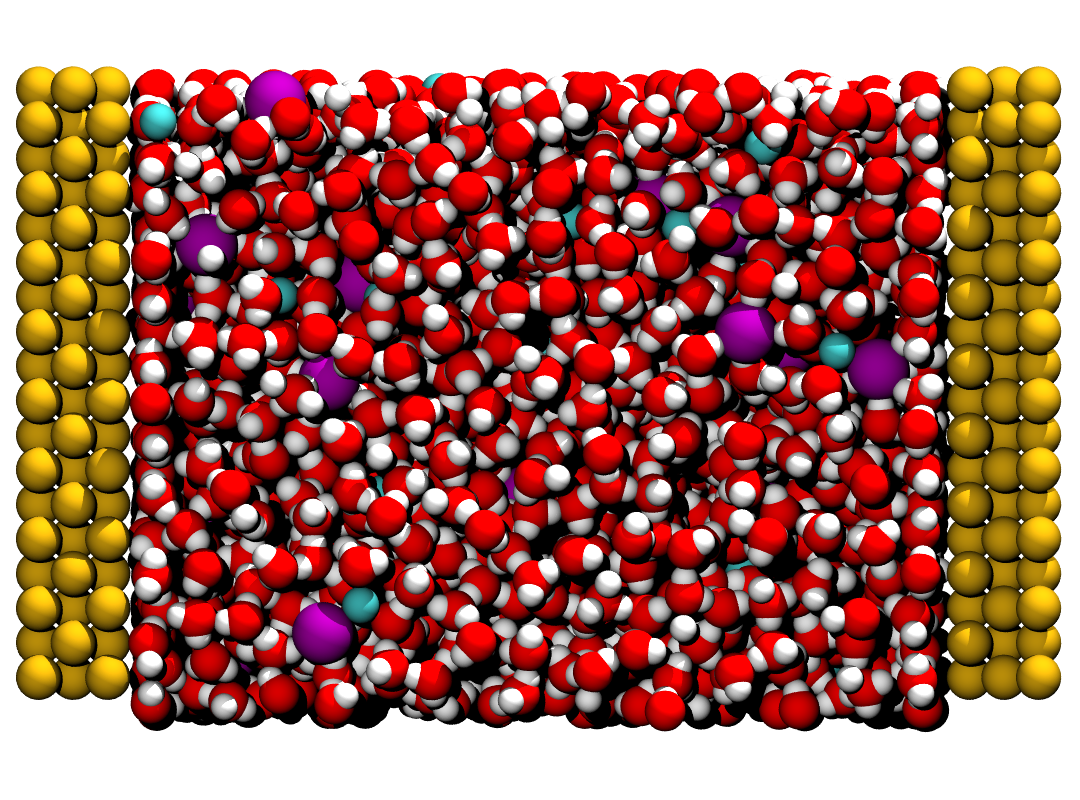}
    \end{minipage}
    \caption{Left: Electric field for three implementations of the constant-$E$ with different integration orders.
    Right: Model system with gold electrodes and aqueous electrolyte.}
    \label{fig:commutation}
\end{figure}

The operators for the electrostat update and the position update commute and their order should not affect the result.
In \cref{fig:commutation} we show this by comparing three implementations of the electrostat with different splittings.
For ``B-ES-ES-A-B'' both steps with $\Delta t/2$ are performed after the update to positions.
For ``B-ES-A-ES-B'' the first update to $E(t) \rightarrow E(t+\Delta t/2)$ is performed before positions are updated.
For ``B-ES-A-B$'$'' an asymmetric integrator is implemented, but with two half step updates for the noise term (\cf{} last line in \cref{eq:half-step-eom}).
This allows for a direct comparison to the two other implementations.
The $E$-field of the trajectories starts to diverge after about \SI{1.5}{\pico\second} as expected for equivalent algorithms.
With no apparent advantage of a symmetric splitting, we have decided to use the asymmetric splitting in \cref{eq:full-eom} for simplicity in our implementation.

\section{Computational Details}
\label{sec:computational}

In this Section, we provide a detailed summary of the computational methodology employed in all simulations performed in this work.
All simulations were performed using the \textsc{Electrode} package~\cite{ahrens-iwers_electrode_2022} of \textsc{Lammps}~\cite{thompson_lammps_2022}, including the new, fully periodic implementation of the electrostat presented in the main text.
We used the electrostat time constant $\tau_0'$ of \SI{100}{\femto\second} in all production simulations discussed below.

\subsection{HCl/Au(100) benchmarks}

The gold electrodes have a (100) orientation and a lattice constant of \SI{4.08}{\angstrom}.
Each electrode has 6 layers and 16 atoms per layer.
The distance between electrodes is \SI{6}{\angstrom}.
The $\eta$ parameter is \SI{1.979}{\angstrom^{-1}}.
The \ce{HCl} molecule has a bond length of \SI{1.274}{\angstrom}, atomic charges of \SI{\pm 0.172}{\elementarycharge} and a resulting dipole of \SI{1.05}{D}.
The charges are calculated for a static system, therefore no force field is required.
The length of the added vacuum of the system in \cref{main-fig:layer-charges}C in the main text is \SI{10}{\angstrom}.

\subsection{[BMIM$^+$][PF$_6^-$]/graphite supercapacitor simulations}

All [BMIM$^+$][PF$_6^-$]/graphite supercapacitor simulations were realized with the Roy--Maroncelli coarse-grained force-field~\cite{roy_improved_2010} to describe the interactions, as adopted from the collection of examples of the  \textsc{Electrode} package~\cite{ahrens-iwers_electrode_2022}.
Similar superacapacitor models have been used in previous studies on \ac{cpm}~\cite{merlet_imidazolium_2011, tee_fully_2022}.
Under this description, each BMIM$^+$ cation is represented as a structured rigid molecule of three bonded beads and each PF$_6^-$ anion is represented as a single spherical bead.
The graphite electrodes are described with Lennard-Jones parameters $\epsilon = \SI{2.38}{\milli\elementarycharge\volt}$ and $\sigma = \SI{3.37}{\angstrom}$~\cite{cole_interaction_1983}.
Our system (as shown in \cref{main-fig:cell-definition} in the main text) consisted of a unit cell with a volume of $\num{31.97}\times\num{34.08}\times\SI{128.26}{\angstrom}^3$, containing a 6-layer graphite slab with 416 atoms per layer (that is, 2496 carbon atoms in total) and 640 BMIM$^+$ and PF$_6^-$ particles each.

\subsubsection{Benchmarks of the dynamics equivalence}

For the purpose of benchmarking the equivalences of the 2D-periodic and 3D-periodic ensembles (\cref{main-fig:equivalence} of the main text), the \acs{md} of the system was evolved under the constant temperature and potential/charge conditions.
The temperature was set to 500~K and was maintained using an explicit Nos\'{e}--Hoover chain thermostat~\cite{nose_unified_1984, hoover_canonical_1985, martyna_nosehoover_1992} using \textsc{Lammps}' \texttt{fix nvt} and an integration time step of 5~fs for equilibrations and 0.5~fs for production.
The latter is an overly conservative choice for this system that only exhibits slow dynamics; we employed it primarily to have a detailed resolution of the resulting time-resolved trajectories.

The constant-potential simulations were performed under a 2~V bias, which was enforced in distinct simulations with all four available approaches: \acs{cpm}, potentiostat, constant-$E$ and electrostat.
For the \acs{cpm} and constant-$E$ method pair, we first performed a 1~ns equilibration with the \acs{cpm} method to fully relax the \acs{edl} and electrode charges under the imposed potential.
We then started the production runs with both the constant-$E$ method and the \acs{cpm} simultaneously from an exactly identical initial condition for 20~ps and monitored the electrode charge $Q$ along both trajectories.
An identical approach was taken for the electrostat and potentiostat pair.
In addition, the random seed in both cases must be fixed in this case so that the sequence of the random numbers for the fluctuation term of the equation of motion was the same to allow for a clear comparison.
The constant-charge simulations targeted an electrode charge of 3~e, which was ensured by the \acs{ccm} method and constant-$D$ method.
The approach taken in this case was identical to the above, only the immediate potential $\phi$ was monitored in this case along the trajectories.

\subsubsection{Diffusion coefficient simulations}
A Nos\'e-Hoover thermostat~\cite{nose_unified_1984, hoover_canonical_1985, martyna_nosehoover_1992} was used with a temperature of \SI{500}{\kelvin}.
The cutoff for pair interactions was set to \SI{15}{\angstrom} and the accuracy parameter for the particle-particle particle-mesh algorithm was set to \num{e-6}.

For each combination of voltage and $E$-field method, the system was first equilibrated for \SI{6}{\nano\second}.
Then, starting points for \num{200} trajectories were created by resetting velocities and then propagating for \SI{2}{\pico\second} for each starting point.
Each of the trajectories was run for \SI{1}{\nano\second} with a time step of \SI{2}{\femto\second}.
Outputs were printed at an interval of \SI{200}{\femto\second}.

In order to calculate diffusion coefficients, the \acp{msd} in $z$ direction were extracted from each trajectory and binned according to their $z$-coordinate with 60 bins between the inner layers of electrodes at \SI{6.7}{\angstrom} and \SI{118.2}{\angstrom}.
The \acp{msd} were calculated as
\begin{equation}
    Z_{I, \zeta}^2(\tau) =\left\langle  \frac{1}{N_{I, \zeta}} \sum_{i=1}^{N_{I, \zeta}} \left[ z_i(\tau) - z_i(0) \right]^2 \right\rangle,
\end{equation}
where $I$ indexes the ion species \ce{PF6-} and \ce{BMIM+} and the sum over $i$ runs over all atoms with $z_i(0) \in [z_{\zeta}, z_{\zeta+1})$, \ie{}, in the respective bin.
The angle brackets indicate a sampling over the possible shifts of $t=0$ within one trajectory.
The center of mass of the three coarse-grained \ce{BMIM+} particles was used as their position, whereas the force-field has only one particle for \ce{PF6-}.
Diffusion coefficients
\begin{equation}
    D_{\perp, I, \zeta} = \frac12 \left\langle \frac{\dd}{\dd \tau} Z^2_{I,\zeta}(\tau) \right\rangle
\end{equation}
were calculated using the Einstein relation \cite{tuckerman_statistical_2010} for each bin by averaging the derivative of the \ac{msd} in a window between \num{200} and \SI{800}{ps}.
Last, the diffusion coefficients of all 200 trajectories were averaged.

\subsection{Aqueous LiCl ion pair separation at Pd(111)}

The interaction potential for the LiCl/water/Pd(111) system (snapshot shown in \cref{main-fig:ions} of the main text) was obtained by combining the INTERFACE~\cite{heinz_thermodynamically_2012} Lennard--Jones parameters for a palladium slab with the rigid SPC/E water model~\cite{berendsen_missing_2002} and Joung--Cheatham ion parameters~\cite{joung_determination_2008} for Li$^+$ and Cl$^-$ using  Lorentz--Berthelot mixing rules~\cite{lorentz_ueber_1881}.
The employed \textsc{Lammps} pair coefficients are summarized in \cref{tab:LiCl-coeff}.
Our system contained 426 water molecules, 1 LiCl ion pair and a 6-layer Pd(111) slab with 8$\times$8 atoms per layer with the surface normal aligned with the $z$-direction.
The volume of the orthorhombic unit cell was $\num{22.2}\times\num{19.226}\times\SI{43.654}{\angstrom}^3$ and was chosen to recreate bulk density of water of 1000~kg~m$^{-3}$ in the middle of the liquid layer.
\begin{table}[tb!]
\begin{center}
    \include{./tables/LiCl_coeff.tex}
    \caption{Lennard-Jones parameter for the Aqueous \ce{LiCl} ion pair at \ce{Pd}.}
    \label{tab:LiCl-coeff}
\end{center}
\end{table}

The dynamics of the system was propagated under the constant-potential ensemble at 300~K with the temperature controlled by an explicit Nos\'{e}--Hoover chain thermostat~\cite{nose_unified_1984, hoover_canonical_1985, martyna_nosehoover_1992} within the \texttt{fix nvt} infrastructure of \textsc{Lammps}, while employing an integration time step of \SI{2}{\femto\second}.
We performed all simulations at potential biases of 0 and 1~V.
We ensured the potential control primarily with our new electrostat implementation and repeated all simulations with the constant-$E$ method for direct comparisons as well as additional validations against the 2D-periodic \acs{cpm} and potentiostat.
Static harmonic biases we imposed using the PLUMED package~\cite{bonomi_promoting_2019, tribello_plumed_2014} coupled to \textsc{Lammps}.
In all simulations, we have kept the Li-cation bound to the surface by imposing a weak harmonic upper-wall restraint on its $z$-coordinate at \SI{4.145}{\angstrom} from the topmost layer of the Pd slab (\ie{}, \SI{11}{\angstrom} from the bottom of the unit cell) with a force constant of \SI{100}{\kilo\joule.\mol^{-1}\angstrom^{-2}} and controlled the ion separation $r$ by a harmonic restraint on the ion-ion distance with a force constant of \SI{500}{\kilo\joule.\mol^{-1}\angstrom^{-2}}.

For both potential control approaches, the electrostat and constant-$E$ method, we first equilibrated each system at the target voltage drops to a total simulation time of 200~ps until the electrode charge has safely converged to an unchanging average value.
During the equilibrations, we enforced the \acs{cip} configuration by restraining $r$ to the value of \SI{2.0}{\angstrom}.

We then directly proceeded to obtain the free-energy profiles along $r$ via the means of an umbrella sampling simulation~\cite{torrie_nonphysical_1977, frenkel_understanding_2002}.
For this purpose, we set up 40 equidistant windows spanning $r$ from \num{2.0} and \SI{7.0}{\angstrom}.
Starting from the equilibrated configuration, we have simulated the windows in a sequential manner, initiating each from the final configuration for the previous one, obtaining \SI{200}{\pico\second} of sampling in each.
We validated a good overlap between the simulated windows in all runs and reconstructed the free-energy profiles from the simulated sampling of $r$ in each window by numerically solving the equations of the binless weighted histogram analysis method~\cite{kumar_weighted_1992} (WHAM) and resolved on a grid of 150 bins.
In a post-processing step, we treated the raw free-energy profiles $F_\mathrm{raw}(r)$ by adding the following configurational entropic (Jacobian) correction 
\begin{equation}
     F(r)
     =
     F_\mathrm{raw}(r) + 2 k_\mathrm{B}T \ln(r),
\end{equation}
which is needed due to the use of a radial collective variable and ensures a plateau at large ion-ion separations.

The error bars of the calculated free-energy profiles were estimated by the block estimation method~\cite{flyvbjerg_error_1989}.
Each window trajectory was split into 20~ps-long contiguous blocks, which were used to compute their corresponding free-energy profiles and the error was obtained as taking the standard deviation of the sample at each $r$ value.
We note that this treatment of error bars puts an upper bound on the true statistical error of our profiles.

\section{Simulation validation}
\label{sec:validation}

\subsection{Temperature control}

The capacitor model from Ref.~\citenum{deisenbeck_dielectric_2021} is simulated with different methods of applying a potential and field.
``Uniform potentiostat'' and ``potentiostat'' use the \ac{ps} as described in Ref.~\citenum{deisenbeck_dielectric_2021} (\cf{} \cref{main-eq:eom-q}) with uniform distribution of charges and distribution using the \ac{cpm}, respectively.
``electrostat'' is using the new constant-$E$ method.
The temperature is not controlled by a thermostat.
The \ac{ps} and constant-$E$ method do not cause drift in the temperature.
The expected noise of the potential $\kb{} T/ C_0$ and the variances in the applied potential are shown in \cref{tab:phi-variances}.

\begin{figure}
    \includegraphics{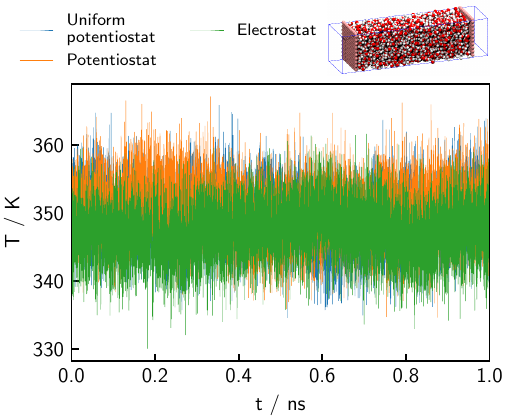}
    \caption{
        Temperature over time of water in a model capacitor.
    }
    \label{fig:temp-control}
\end{figure}

\begin{table}[H]
\begin{center}
    \include{./tables/deissenbeck_variances.tex}
    \caption{
        Temperature, vacuum capacitance as well as expected and observed variance of the voltage for a model system (\cf{} inset of \cref{fig:temp-control}).
        The expected variance $\kb{} T / C_0$ is given by the Johnson--Nyquist theorem~\cite{blundell_concepts_2017}.
        For the \ac{ps} and constant-$E$, $C_0$ was determined from the $\mathbf A$-matrix (\cf{} \cref{eqn:C0}).
        For the uniform \ac{ps}, the vacuum capacitance was estimated with the equation for an ideal plate capacitor.
        The voltage $\phi$ was extracted from \SI{1}{\nano\second} trajectories wit a time step of \SI{1}{\femto\second}.
    }
    \label{tab:phi-variances}
\end{center}
\end{table}

\subsection{Benchmark of ensemble equivalence}
\begin{figure}[tb!]
    \includegraphics{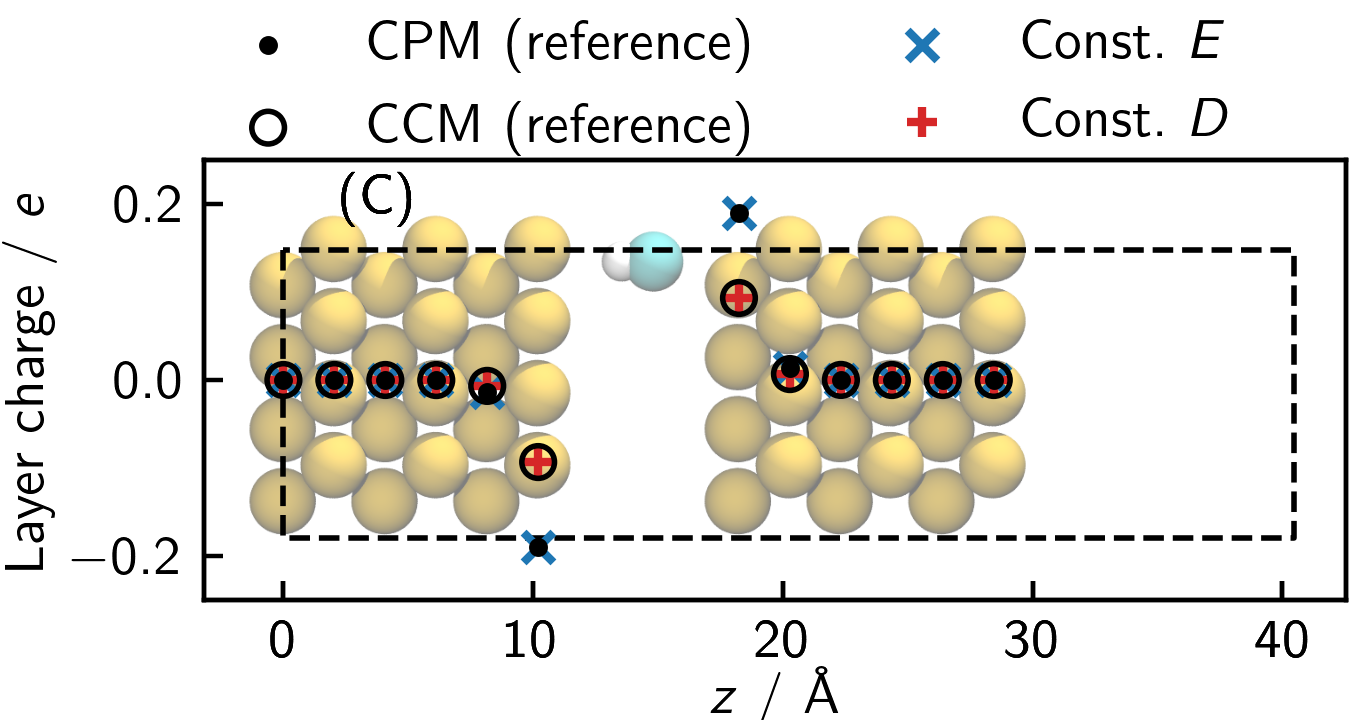}
    \caption{
    Equivalences of different electrostatic descriptions of a model HCl/Au(100) nanocapacitor.
    The model is identical to the one in \cref{main-fig:layer-charges}A except for the added vacuum in $z$-direction.
    The unit cell is indicated with dashed lines, showing the increased length in $z$.}
    \label{fig:benchmark-vacuum}
\end{figure}

The converged results for a model with 2D \acp{pbc} are not affected by additional vacuum outside of the model system in the non-periodic direction.
Even though the constant-$E$ and -$D$ methods use 3D \acp{pbc}, additional vacuum should not affect the result either.
We show this for the benchmark model from \cref{main-fig:layer-charges} by adding vacuum to the right in the $z$-direction in \cref{fig:benchmark-vacuum}.
The equivalences between \ac{cpm} and constant-$E$, and between \ac{ccm} and constant-$D$ still hold and the values for each layer are the same as for the model without vacuum (\cf{} \cref{tab:charges-per-layer}).

The constant-$D$ calculations in \cref{tab:charges-per-layer} use $D^* =\SI{0.1}{\elementarycharge} / A$, scaled by $g=L'/L$ (\cf{} \cref{main-eqn:eff-D}).
Instead of using the distance between $z$-coordinates of atoms in the inner layers of electrodes, we calculate an effective distance $L' = \epsilon_0 A / C_0$, where $C_0$ is obtained from \cref{eqn:C0}.
This form accounts for the fact that the electrode is comprised of atoms with Gaussian shaped charge distributions, which do not exactly reproduce the charge distribution of an ideal plate capacitor.
In this case, the values for the plate-to-plate distance and the effective distance are \SI{8.04}{\angstrom} and \SI{8.320}{\angstrom}, respectively.
\begin{table}[H]
\begin{center}
    \include{./tables/layer_charges.tex}
    \caption{Charges of entire right electrode and first three inner layers for every tested setup (\cf{} \cref{main-fig:layer-charges}).
    The ``System'' column refers to the geometry shown in the respective panel in \cref{main-fig:layer-charges,fig:benchmark-vacuum}.}
    \label{tab:charges-per-layer}
\end{center}
\end{table}

\section{Additional results}
\label{sec:additional-results}

\subsection{Total current $j$ vs electrolyte current $j_\mathcal{E}$}

\begin{figure}[tb!]
    \includegraphics[width=0.75\linewidth]{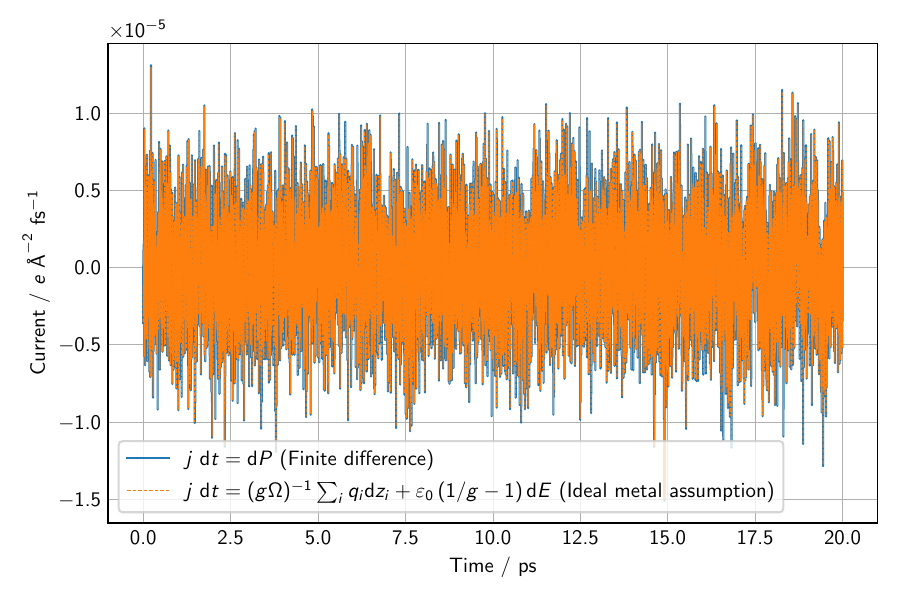}
    \caption{
    The evolution of the total current along a electrostatted trajectory of the [BMIM$^+$][PF$_6^-$]/graphite ionic liquid supercapactor.
    Blue: the total current obtained from the finite difference of the total polarization, $j = \dd P / \dd t$.
    Orange: the total current calculated on the premise of an ideal metal electrode, which uses only the electrolyte current $j_\mathcal{E}$ and the stochastic $E(t)$.
    }
    \label{fig:current}
\end{figure}

Here, we assess the validity of the ideal metal assumption for the \ac{cpm}-based metal embedded into our \textsc{lammps} electrostat framework.
We measure \textit{a posteriori} the total current alongside a electrostatted trajectory of the ionic liquid supercapacitor discussed in the main text.
Specifically, we do so in two distinct ways.
First, we use the definition of $j$ to compute its trajectory as a finite-difference derivative of the total polarization
\begin{equation}
    j
    =
    \frac{\dd P}{\dd t}
\end{equation}
with full resolution.
Second, we rely on \cref{main-eq:linear-polarization} of the main text to calculate
\begin{equation}
    j
    =
    \frac{1}{g} \frac{\dd P_\mathcal{E}}{\dd t} + \varepsilon_0 \left(\frac{1}{g} - 1\right) \frac{\dd E}{\dd t}
    =
    \frac{1}{g\Omega} \sum_{i \in \mathcal{E}} q_i\dot{z}_i + \varepsilon_0 \left(\frac{1}{g} - 1\right) \frac{\dd E}{\dd t};
\end{equation}
we take the time derivatives of stochastic quantities symbolically, as they are not, strictly speaking, well defined due to the discontinuity of $E(t)$.
The two trajectories are shown in \cref{fig:current} and they exhibit a numerically identical behavior.
It is a consequence of the ideal-metal assumption in the \ac{cpm} (and all the related methods).

\subsection{Ion association as described by \ac{cpm} and \ac{ps}}

\begin{figure}[tb!]
    \includegraphics{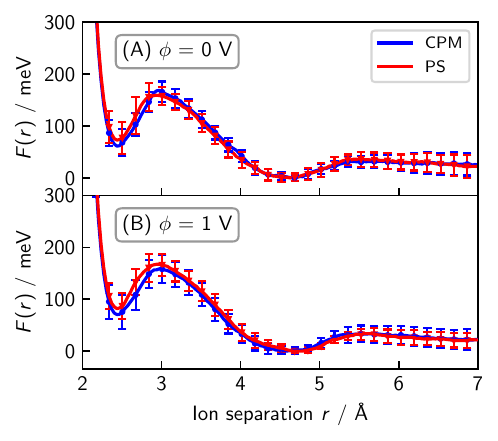}
    \caption{
    The same ion-association free-energy profiles as presented in \cref{main-fig:ions} of the main text, just calculated with the \ac{cpm} and \ac{ps} methods and under 2D \ac{pbc}.
    In this case, the error is shown as error bars rather than shaded intervals.
    }
    \label{fig:ions-cpm-vs-ps}
\end{figure}

In \cref{fig:ions-cpm-vs-ps}, we present the ion-association free-energy profiles complementary to the ones presented in \cref{main-fig:ions} of the main text, calculated with \ac{cpm} and with \ac{ps}.
Consistently with the equivalence of the potential-based and field-based methods, the profiles show exactly the same characteristics as the ones in the main text.
We have performed independent equilibrations also in this case and, within the estimated error bars, this points to the fact that the minor, but systematically appearing deviation seen the main text is of statistical origin.

\subsection{Charge dynamics}

\begin{figure}
    \includegraphics[width=0.5\linewidth]{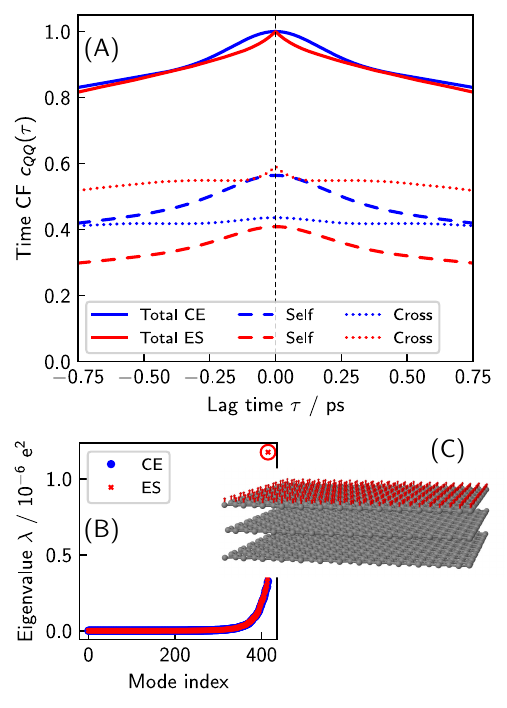}
    \caption{
    Analysis of the charge dynamics of the electrode.
    Panel A shows the autocorrelation function $c_{QQ}(\tau)$ (solid) for the constant-$E$ method (blue, ``CE'') and electrostat (red, ``ES'') simulations of the ionic liquid supercapacitor (\cf{} Section~\ref{main-sec:diffusion} of the main text and Section~\ref{sec:computational} for details) and its decomposition into self- (dashed) and cross-term (dotted) components.
    Note the smooth, zero-derivative onset in all of the blue curves and the exponential cusp in the red total and cross-term curves at $\tau = 0$.
    Panel B shows the eigenvalue spectrum of the short-time covariance matrix of the atomic charges.
    Note the the noticeably separated highest eigenvalue of the electrostat simulation marked by a red circle.
    Panel C shows a visual representation of the perfectly uniform eigenvector corresponding to the marked eigenvalue on a snapshot of the graphite electrode.
    The eigenvector belongs to the vector space of atomic charges: we use the arrow direction to denote the sign of charging of each atom and its relative magnitude to denote its extent (the absolute magnitude has no meaning).
    }
    \label{fig:acfs}
\end{figure}

\begin{figure}[tb!]
    \includegraphics{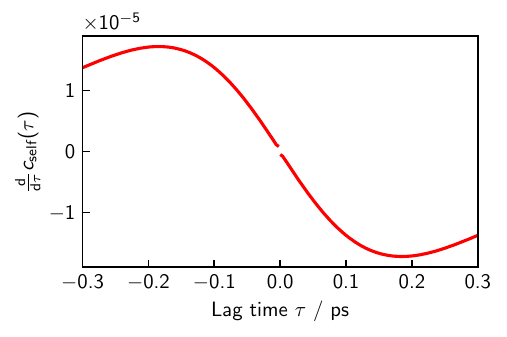}
    \caption{
    Time derivative of the constant-$E$ \ac{acf} $\dd c_\mathrm{self}(\tau) / \dd\tau$ manifests a minor discontinuity around $\tau = 0$.
    }
    \label{fig:acf-self-dt}
\end{figure}

In the following, we provide a complementary analysis to Section~\ref{main-sec:size-effect} of the main text on the relevance of electrostat fluctuations.
As discussed in \cref{main-fig:equivalence} of the main text, the electrode charge dynamics under constant-$E$ conditions is fully deterministic and is given by the time evolution of the electrolyte, while under the electrostat the charge dynamics contain a high-frequency noise component due to the stochastic term, which can be quantified through time-correlation functions.
The \acf{acf} of the total electrode charge $Q$
\begin{equation}
    c_{QQ}(\tau)
    =
    \frac{\langle \delta Q(t) \delta Q(t + \tau) \rangle}{\langle \delta Q^2 \rangle},
\end{equation}
where $\delta Q(t) \equiv Q(t) - \langle Q \rangle$,
will be smooth and differentiable under constant-$E$ conditions with $\dot{c}_{QQ}(0) = 0$ and a parabolic onset for short $\tau$ (we note that here, we are using a Nos\'{e}--Hoover-chain thermostat to control the temperature of the electrolyte; the situation might change if one employed a stochastic thermostat).
When the charge dynamics is made stochastic by the electrostat, the injected non-differentiable noise manifests as a spike at a zero lag and an exponential onset of the \ac{acf} proportional to $e^{-|\tau|/\tau'_0}$~\cite{blundell_concepts_2017}.
This exact behavior is captured in \cref{fig:acfs}A (solid curves).
Given that $Q(t) = \sum_{i \in \mathrm{electrode}} q_i(t)$, we can decompose the \ac{acf} into a self term (which addresses the time correlation of an atomic charge with itself) and a cross term (which correlates charges on different atoms),
\begin{equation}
    c_{QQ}(\tau)
    =
    \sum_i c_{q_i q_i}(\tau) 
    +
    \sum_{i \neq j} c_{q_i q_j}(\tau)
    \equiv
    c_{QQ}^{\mathrm{self}}(\tau)
    +
    c_{QQ}^{\mathrm{cross}}(\tau).
\end{equation}
This decomposition is also shown in \cref{fig:acfs}A.
In the deterministic constant-$E$ case, both terms clearly have a flat onset  but the cusp in the electrostat case is only appreciably present in $c_{QQ}^{\mathrm{cross}}(\tau)$ while the $c_{QQ}^{\mathrm{self}}(\tau)$ exhibits nearly perfectly smooth dynamics (a minute discontinuity, only visible in the time derivative of the self term, is presented in \cref{fig:acf-self-dt}).
This result offers an intriguing perspective: the individual charge dynamics is effectively void of noise, which hides in cross-correlations and, consequently, the electrolyte dynamics is affected only little.

The reason behind this effect is simple and can be attributed to the fact that the electrostat noise gets transferred onto the atoms through a spatially uniform charge mode.
To get this insight, we performed an eigendecomposition of the covariance matrix of the atomic charge dynamics, after applying a high-pass filter beforehand. 
Specifically, for each atomic charge time series, we calculated the following difference
\begin{equation}
    \delta q_i(t)
    =
    q_i(t + \delta t) - q_i(t),
\end{equation}
which is equivalent to a simple linear Fourier filter that favors high  (here, we have chosen the value of $\delta t$ = \SI{10}{\femto\second}; the results obtained are similar up to \SI{50}{\femto\second}, and at higher values they start to be affected by the lower frequencies of the physical electrolyte dynamics).
With the filtered time series, we build a covariance matrix
\begin{equation}
    K_{ij}
    =
    \mathrm{cov}(\delta q_i, \delta q_j),
\end{equation}
which we diagonalized to find the set of eigenmodes $\{\mathbf{a}_i\}$ and eigenvalues $\{\lambda_i\}$ such that
\begin{equation}
    \mathbf{K} \mathbf{a}_i = \lambda_i \mathbf{a}_i.
\end{equation}
The eigenvalue spectrum is shown in Fig.~\ref{fig:acfs}B.
This procedure allowed us to obtain a set of charge modes and their corresponding eigenvalues that correspond to the variance of each mode within the fast dynamics.
The modes with the highest variance are of most interest.
The highest eigenvalue mode in the electrostat simulation stands out particularly strongly (note that inset snapshot does not hide any points, the highest electrostat eigenvalue is pronouncedly separated).
This mode, visualized in \cref{fig:acfs}C where werely on a real-space visualization using spatial arrows to convey an impression of the relative magnitude and ``direction'' of the changing atomic charge, is a uniform charging mode.

Therefore, the noise is injected into each atom dominantly with the same magnitude.
If we let $N$ denote the number of atoms in the contact layer of the electrode then, specifically, for the simulated [BMIM$^+$][PF$^-_6$]/graphite system, $N = 416$.
As a simple model, we can decompose each atomic charge time series as
\begin{equation}
    q_i(t)
    =
    s_i(t) 
    +
    \eta(t),
\end{equation}
where $s_i$ represents a smooth, deterministic part and $\eta$ is a random noise variable of zero mean and appropriate variance (which is identical for every $i$).
In the language of \acp{acf}, the self terms then become
\begin{equation}
    c^\mathrm{self}_{QQ}
    =
    \sum_{i=1}^N c_{s_i s_i}
    +
    N c_{\eta \eta}
\end{equation}
while the cross terms are
\begin{equation}
    c^\mathrm{cross}_{QQ}
    =
    \sum_{i \neq j}^N c_{s_i s_j}
    +
    N(N - 1) c_{\eta \eta}.    
\end{equation}
Clearly, the cross term gets $(N - 1)$-times the intensity of the noise \ac{acf} than the self term, which explains the earlier lack of the cusp in the electrostat $c_{QQ}^\mathrm{self}(\tau)$.
Under these considerations, we conclude that the forces on the atoms of the electrolyte, which drive its dynamics and which are governed by the  Coulomb law and thus feel contributions mainly from the nearest neighboring atoms, could feel significant noise only with a much smaller $N$ than the one considered in Section~\ref{main-sec:diffusion}.
This conclusion is consistent with the corresponding results presented in the main text.

\bibliographystyle{unsrt}
\bibliography{main}

%% file: acronyms.tex
\begin{acronym}[LAMMPS]
    \acro{ccm}[CCM]{constrained charge method}
    \acro{cpm}[CPM]{constant potential method}
    \acro{ps}[PS]{potentiostat}
    \acro{es}[ES]{electrostat}
    \acro{edl}[EDL]{electric double layer}
    \acro{ssv}[SSV]{Stengel--Spaldin--Vanderbilt}
    \acro{sde}[SDE]{stochastic differential equation}
    \acro{pbc}[PBC]{periodic boundary condition}
    \acro{md}[MD]{molecular dynamics}
    \acro{acf}[ACF]{auto-correlation function}
    \acro{cip}[CIP]{contact ion pair}
    \acro{ssp}[SSP]{solvent-separated pair}
    \acro{msd}[MSD]{mean square displacement}
\end{acronym}

%% file: tables/LiCl_coeff.tex
\begin{tabular}{l|l|l}
Element & $\epsilon / \unit{\kilo cal\,\mol^{-1}}$ & $\sigma / \unit{\angstrom}$\\\hline
Pd & 6.15 & 2.5114\\
H & 0 & 0\\
O & 0.1553 & 3.166\\
Li & 0.3367 & 1.409\\
Cl & 0.0128 & 4.83\\
\end{tabular}

%% file: tables/deissenbeck_variances.tex
\begin{tabular}{l|r|r|r|r}
Method & $T / \unit{K}$ & $C_0 \;/\; \unit{eV^{-1}}$ & $\kb{}T/C_0 \;/\; \unit{V^2}$ & $\langle\phi^2\rangle \;/\; \unit{V^2}$\\\hline
Uniform Potentiostat & 350 & 0.0432 & 0.699 & 0.695\\
Potentiostat & 350 & 0.0433 & 0.696 & 0.713\\
Electrostat & 350 & 0.0440 & 0.685 & 0.676\\
\end{tabular}

%% file: tables/layer_charges.tex
\begin{tabular}{l|l|l|l|l|l}
Method & System & $q_\mathrm{all} / e$ & $q_1 / e$ & $q_2 / e$ & $q_3 / e$\\\hline
CPM (reference) & (A) & \num{2.0327e-01} & \num{1.8946e-01} & \num{1.3650e-02} & \num{1.9321e-04}\\
const.-$E$ & (A) & \num{2.0327e-01} & \num{1.8946e-01} & \num{1.3654e-02} & \num{1.9063e-04}\\
const.-$E$ & (B) & \num{2.0327e-01} & \num{1.8946e-01} & \num{1.3654e-02} & \num{1.9063e-04}\\
const.-$E$ & (C) & \num{2.0327e-01} & \num{1.8946e-01} & \num{1.3654e-02} & \num{1.9064e-04}\\
CCM (reference) & (A) & \num{1.0000e-01} & \num{9.3199e-02} & \num{6.7233e-03} & \num{9.6184e-05}\\
const.-$D$ & (A) & \num{1.0001e-01} & \num{9.3202e-02} & \num{6.7262e-03} & \num{9.4331e-05}\\
const.-$D$ & (B) & \num{1.0001e-01} & \num{9.3202e-02} & \num{6.7262e-03} & \num{9.4331e-05}\\
const.-$D$ & (C) & \num{1.0001e-01} & \num{9.3202e-02} & \num{6.7262e-03} & \num{9.4333e-05}\\
\end{tabular}